\documentclass[useAMS,usenatbib,usegraphicx]{mn2e}

\usepackage{amsmath, hyperref, amssymb, color, graphicx, latexsym}
\usepackage{multirow}
\usepackage{xspace}

\newcommand{\gws}{gravitational waves\xspace}
\newcommand{\gw}{gravitational wave\xspace}
\newcommand{\eos}{EoS\xspace}
\newcommand{\eoss}{EoS\xspace}
\newcommand{\ee}[1]{\!\times\!10^{#1}}

\title{Prospects of observing continuous \gws from known pulsars}

\author[M. Pitkin]{Matthew Pitkin$^1$\thanks{matthew.pitkin@glasgow.ac.uk} \\
$^1$SUPA, School of Physics and Astronomy, University of Glasgow,
University Avenue, Glasgow, G12 8QQ, UK}
\date{}

\pagerange{\pageref{firstpage}--\pageref{lastpage}} \pubyear{2011}

\begin{document}

\maketitle

\begin{abstract}
Several past searches for \gws from a selection of known pulsars have been
performed with data from the science runs of the Laser Inferometer
Gravitational-wave Observatory (LIGO) \gw detectors. So far these have lead to
no detection, but upper limits on the \gw amplitudes have been set. Here
we study our intrinsic ability to detect, and estimate the \gw amplitude for
non-accreting pulsars. Using spin-down limits on emission as a guide we examine
amplitudes that would be required to observe known pulsars with future detectors
(Advanced LIGO, Advanced Virgo and the Einstein Telescope), assuming that they
are triaxial stars emitting at precisely twice the known rotation frequency.
Maximum allowed amplitudes depend on the stars' equation of state (e.g.\ a
normal neutron star, a quark star, a hybrid star) and the theoretical mass
quadrupoles that they can sustain. We study what range of quadrupoles, and
therefore \eoss, would be consistent with being able to detect these sources.
For globular cluster pulsars, with spin-downs masked by accelerations within the
cluster, we examine what spin-down values \gw observations would be able to set.
For all pulsars we also alternatively examine what internal magnetic fields they
would need to sustain observable ellipticities.
\end{abstract}


\section{Introduction}\label{sec:intro}
Just under 2000 pulsars are currently known \citep{Manchester:2005} and these
provide enticing targets to search for continuous \gw emission. Searches for
\gws from known pulsars \citep{Abbott:2005, Abbott:2007a, Abbott:2010a} estimate
four unknown parameters: the \gw amplitude $h_0$, the initial phase relative to
the electromagnetic pulse phase $\phi_0$, the cosine of the orientation angle
$\cos{\iota}$, and the polarisation angle $\psi$. As yet no signal has been
seen from any pulsar, so observations provide no constraints on these parameters
other than an upper limit on $h_0$. However, with detector sensitivities
improving over the next few years we will be beating spin-down based upper
limits for many pulsars. This gives us the potential of direct \gw detection
(if \gws provide a braking mechanism that is comparable to the electromagnetic
braking) therefore allowing us to place real constraints on these parameters.
The $h_0$ parameter in particular will allow us to say something about the make
up of the neutron star\footnote{We will use `neutron star' as a generic term for
all types of compact star used in this paper, unless specifically stated.}
itself as it is
related to the star's mass quadrupole moment, and therefore the equation of
state (EoS). Some discussion of useful astrophysics that could be gained from
measuring $\cos{\iota}$ is given in \citet{Jones:2007}, and accurate
polarisation measurements could provide limits on, or allow the study of,
different theories of gravity (see e.g.\ \citealp{Will:2006}), but here we will
concentrate on the amplitude and quadrupole measurement. It is important to
stress that there is much that is unknown about neutron stars and 
whether they can produce and sustain large enough quadrupoles to be observable
via gravitational waves, so explaining such issues is still an area of great
interest. In this study we only discuss prospects for observing
non-accreting pulsars. In accreting systems different \gw emission mechanism may
be present, and the links between electromagnetic observations and any
gravitational wave signal are less well known. \citet{Watts:2008} provide a good
study of detection prospects for accreting neutron stars.

\subsection{Searches for known pulsars}
Since the start of science data taking from the current generation of
interferometric \gw detectors (LIGO, GEO\,600, Virgo and TAMA
\citealt{Abbott:2004, Acernese:2008, Ando:2005}) searches looking for \gws from
a large selection of {\it known} pulsars (millisecond and young pulsars with
spin frequencies greater than $\sim 20$\,Hz) have been performed
\citep{Abbott:2005, Abbott:2007a, Abbott:2010a}. In the most recent analysis 116
pulsars were searched for using approximately a year and a half of data from
each of the three LIGO detectors \citep{Abbott:2010a}, which were operating at
their design sensitivity \citep{Abbott:2009a}. Unfortunately no signal was seen
from any of these objects, but for the majority the sensitivity was still well
above, by factors of 10 to over 100 times, their spin-down limits. The spin-down
limit is set by assuming that the star's spin-down luminosity is equal to its
\gw luminosity i.e.\ all the rotational energy lost by the pulsar is due to
radiation via gravitational waves from the $Q_{22}$ mass quadrupole. This limit
does require one to assume a moment of inertia for the star, generally taken as
the canonical value of $10^{38}$\,kg\,m$^2$ (or $10^{45}$\,g\,cm$^2$ in cgs
units), and that its distance is precisely known. For one object, the Crab
pulsar, this limit has been passed \citep{Abbott:2008a, Abbott:2010a} although
still no \gws were seen, and for four others the upper limit obtained was within
a factor of 10 from spin-down.

For known pulsars the parameter space to be searched over is comparatively
small (position and phase evolution are known) and long observation times can be
used in a coherent way. This makes such searches more sensitive than
semi-targeted, or blind searches for similar sources \citep{Abbott:2007b,
Abbott:2008b, Abbott:2009b, Abbott:2009c}, although potentially will miss out on
some interesting, but currently unknown, objects. \citet{Knispel:2008}
discusses the potential strength of a population of \gw emitting Galactic
neutron stars. Here, we therefore concentrate our study on estimating the
prospects for fully targeted searches. 

The paper is set out as follows: in \S\ref{sec:snrestimates} we will assess the
potential signal-to-noise ratios at which currently known pulsars could be
observed with future detectors, review a detection statistic for these pulsars, and
demonstrate the parameter estimation capabilities of the standard search
technique at a variety of signal-to-noise ratios; in \S\ref{sec:eos} we review
some estimates of the maximum quadrupole moments for neutron stars given a
selection of \eoss; in \S\ref{sec:limits} we assess the potential
signals, and associated quadrupoles, observable in future detectors based on
spin-down limits and limits on our sensitivity, and how these limits can be
thought of in terms of internal magnetic field strengths, and for globular
cluster pulsars limits on their spin-down. Parts of this work are similar in
scope to the review by \citet{Owen:2006}, and the discussions in
\citet{Abbott:2007b} and \citet{Andersson:2009}.

\section{Estimating signals in future detectors}\label{sec:snrestimates}
\begin{figure}
\includegraphics[width=84mm]{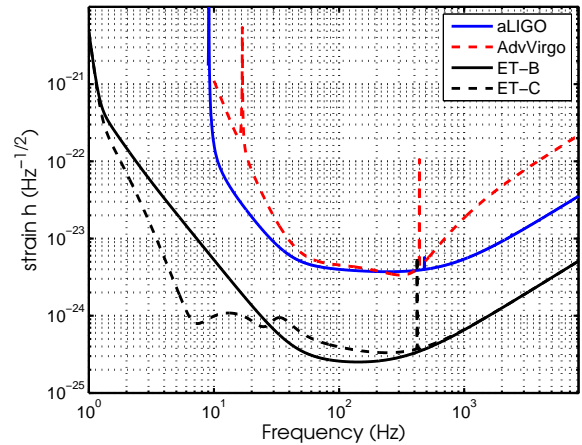}
\caption{Design strain curves for Advanced LIGO (aLIGO) \citep{LSC:2009}
(using the zero detuning, high laser power configuration), Advanced Virgo
(AdvVirgo) (the baseline sensitivity used here is that of 12 May 2009 found at
\url{http://wwwcascina.virgo.infn.it/advirgo/}), and the Einstein Telescope (ET)
in both ET-B and ET-C configurations (taken from 
\url{http://www.et-gw.eu/etsensitivities}).}
\label{fig:strains}
\end{figure}

The next (2$^{\rm nd}$) generation of interferometric \gw detectors, such as
Advanced LIGO (aLIGO) \citep{Harry:2010}, Advanced Virgo (AdvVirgo)
\citep{Virgo:2009}, the Large-scale Cryongenic Gravitational wave Telescope
(LCGT) \citep{Kuroda:2010} and GEO-HF \citep{Willke:2006}, expect to provide
order of magnitude sensitivity improvements over current detectors, and offer
the opportunity to beat spin-down limits for nearly 60 pulsars (see
\S\ref{sec:second}). A 3$^{\rm rd}$ generation \gw detector called the Einstein
Telescope (ET) \citep{Punturo:2010} is also under design study, and would offer
another order of magnitude increase in sensitivity. This would bring hundreds of
currently known pulsars into the range where we could beat spin-down limits (see
\S\ref{sec:third}), and also may coincide with the completion of the Square
Kilometre Array (SKA) radio telescope, which may give us a vastly larger number
of sources to target. Estimates suggest the SKA may detect over half of the
observable pulsars within the Galaxy giving $\sim 20\,000$ potential sources
\citep{Cordes:2004} with $\sim 1000$ of them being millisecond pulsars and some
of which could have large spin-down luminosities. Design strain curves for
aLIGO, AdvVirgo and the ET in two different potential configurations are shown
in Fig.~\ref{fig:strains}.

Assuming that the star is emitting at its spin-down limit, and that a fully
coherent search can be performed, we have estimated the angle averaged
signal-to-noise ratios (S/N) for all known pulsars for which a spin-down limit
can be calculated\footnote{Many of the millisecond pulsars found so far are seen
in globular clusters, and accelerations within these clusters can mask the
spin-down and even produce an observed spin-up, so in this section we exclude
these.} given one year observation times, for joint aLIGO and AdvVirgo
observations (ALV) and for the ET in two potential configurations - ET-B and
ET-C (see \S\ref{sec:third}).

For the Crab pulsar we will estimate the S/N using a limit based on the observed
upper limit, which at $h_0^{95~{\rm per~cent}} = 2.0\ee{-25}$
\citep{Abbott:2010a} is approximately 7 times below the spin-down limit of
$1.4\ee{-24}$ for a star with the canonical moment of inertia and a distance of
2\,kpc. This result also beats limits of \citet{Palomba:2000} based on the Crab
pulsar's observed braking index. For the Vela pulsar and B0540$-$69
(J0540$-$6919), for which braking indices can also be reliably measured, we use
upper limit estimates based on those calculated by \citet{Palomba:2000}. For
Vela (using the Vela age of 11\,000 years given in \citealp{Abadie:2011}) this
gives limits 4 times lower than the spin-down limit of $3.3\ee{-24}$, and for
B0540$-$69 this gives a limit 6 times lower than the spin-down limit of
$5.2\ee{-26}$. For these three pulsars our S/N estimates used known values of
the pulsar inclination angle and \gw polarisation angles as taken from fits to
the pulsar wind nebulae in \citet{Ng:2008}. As can be seen in
Fig.~\ref{fig:AL_ET_SNRs} these three pulsars would have S/N greater than 5 for
all future detectors if emitting at their spin-down, or observationally
constrained, upper limits. They are not included in the numbers quoted in the
rest of this section.

\subsection{Assessing detection and parameter estimation}\label{sec:assess}
We can estimate the S/N for all known pulsars for a given \gw detector, or set
of detectors, to assess our ability detect them and perform parameter
estimation. The S/N, $\rho$, can be calculated from the square root of the inner
product of the signal $h$ with itself, $\rho = \sqrt{(h|h)}$, (the inner, or
scalar, product for real time domain functions is given by
$(x|y) = (2/S_n(\nu))\int_0^T x(t)y(t) {\rm d}t$) which given the time domain
signal model of (e.g.\ \citealp{Dupuis:2005})
\begin{eqnarray}\label{eq:signal}
h(t) & = &  h_0\bigg[\frac{1}{2}(1 + \cos{}^2\iota)F_+(t,\psi)\cos{\phi(t)} +
\nonumber \\
 & & \cos{\iota}F_{\times}(t,\psi)\sin{\phi(t)}\bigg],
\end{eqnarray}
gives
\begin{eqnarray}\label{eq:snr}
\rho & = & \bigg[\frac{h_0^2}{S_n(\nu)}\frac{T}{N} \sum_k^N \Big( [
\frac{1}{2}(1 + \cos{}^2\iota)F_+(t_k,\psi) ]^2 \nonumber \\ 
& & + [ \cos{\iota} F_{\times}(t_k,\psi)]^2\Big)\bigg]^{1/2},
\end{eqnarray}
under the assumption that $\nu T \gg 1$, and where $S_n(\nu)$ is the frequency
dependent one sided noise power spectral density\footnote{This is related to the noise
variance via $S_n = 2\sigma^2 \Delta{}t$, where $\Delta{}t$ is the sample interval
in the data time series.} 
in Hz, $T$ is the total observation
time in seconds, $N$ is the number of samples used, and $F_+$ and $F_{\times}$ are the
source position and polarisation dependent detector antenna patterns (see
e.g.\ \citealt{Jaranowski:1998}). 

The S/N depends on the orientation of the pulsar, so upper and lower values can
be set for the best and worst case orientations. The best case in terms of S/N
is for a pulsar's spin axis to be along the line of sight ($\iota =
0^{\circ}$ or $180^{\circ}$, or $\cos{\iota} = \pm 1$), which gives rise to
circularly polarised radiation; the worst case is for the pulsar's spin axis to
perpendicular to the line of site ($\iota = 90^{\circ}$ or $\cos{\iota} = 0$),
which gives rise to just linearly polarised radiation. The S/N ranges over a
factor of $\sim 3$ between best and worst cases, and the angle averaged, or
expected, S/N (averaging over a uniform distribution in $\cos{\iota}$, $\psi$
and $\phi_0$) is $\sim 1.69$ times below that for best case orientation value.
For a fixed S/N the angle averaged value of $h_0$ is 1.89 times the value needed
to give the same S/N for the best case orientation\footnote{Note that the angle
averaged S/N for a given value of $h_0$ and the angle averaged $h_0$ for a given
S/N are not interchangeable. This comes about because for the first case the
angle averaged S/N $\langle \rho \rangle \propto h_0\langle C \rangle$, whereas
in the second case the angle averaged amplitude $\langle h_0 \rangle \propto
\rho\langle 1/C \rangle$, where $C = \sum_k^N ( [ \frac{1}{2}(1 +
\cos{}^2\iota)F_+(t_k,\psi) ]^2 + [ \cos{\iota}
F_{\times}(t_k,\psi)]^2)^{-1/2}$ and $\langle f(y) \rangle = \int_{y_{\rm
min}}^{y_{\rm max}} p(y) f(y) {\rm d}y$. For an individual pulsar, i.e.\ a
single sky position, the dependence on polarisation angle disappears for data
spans a lot greater than a day, and as can be seen from Eqn.~\ref{eq:snr} the
phase dependence is no longer present, so the orientation angle is the only one
needing averaged over for which we use a uniform distribution between $-1$ and 1
giving $p(\cos{\iota}) = 1/2$.} (i.e.\ for the best case orientation a smaller
value of $h_0$ will produce an equivalent S/N). In this section we
will be asking the question ``Given that the star is emitting \gws at an
amplitude X, what is the expected S/N it would have?'', whereas in
\S\ref{sec:limits} we will be asking the question ``Given that the signal has an
S/N of X, what will the expected $h_0$ (or parameter such as ratio to spin-down
limit or mass quadrupole that scales with $h_0$) be?''. When estimating the
number of pulsars that are potentially observable using either an expected S/N
calculated from a given $h_0$, or an expected $h_0$ calculated from a given S/N,
then the results will be slightly different.

\subsubsection{Detection statistic}
For the rest of this paper we will assume an S/N for a signal that we believe
provides a good chance of detection. To do this we will use Bayesian hypothesis
testing to provide a detection statistic with which to assess detection
efficiency at different S/N (see e.g.\ \citealp{Clark:2007} or
\citealp{Prix:2009}). We produced a Bayes factor in which the two competing
models are that of data containing Gaussian noise {\it and} a known pulsar
signal compared to data just containing Gaussian noise. Uniform priors were set
for the signal parameters, and a prior odds for the competing models of 1 was
used. To assess this we produced Monte Carlo simulations of 2000 signals, with
fixed sky position, but randomly chosen values of $\cos{\iota}$, $\psi$ and
$\phi_0$, in different realisations of noise at a range of S/N. The approach is
essentially using what is described as the $\mathcal{B}$-statistic in
\citet{Prix:2009}, which they show to be slightly more efficient, although
computationally more expensive, than the more widely used 
$\mathcal{F}$-statistic \citep{Jaranowski:1998, Abbott:2004b}.

\begin{figure}
\begin{tabular}{c}
\includegraphics[width=65mm]{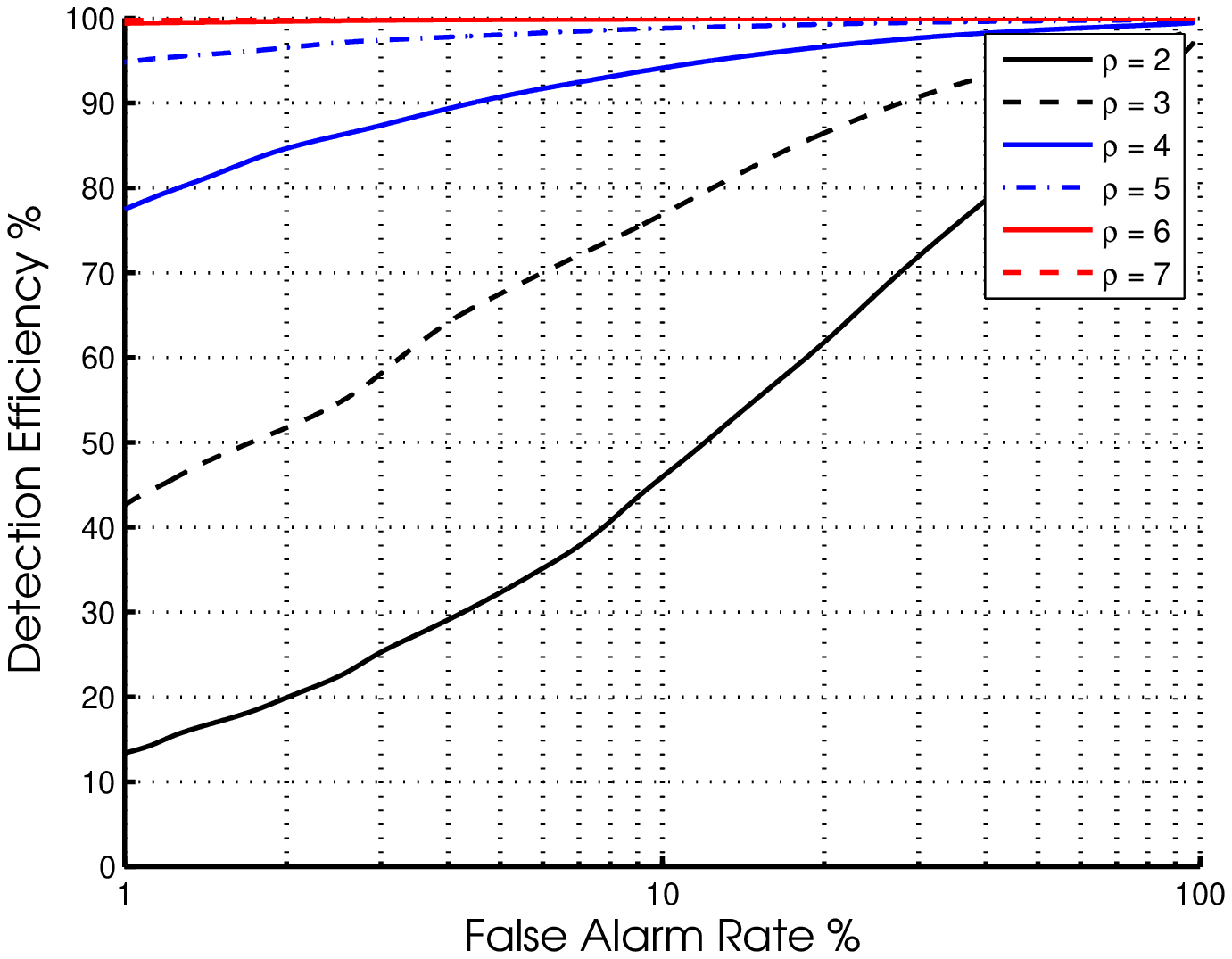} \\
\includegraphics[width=65mm]{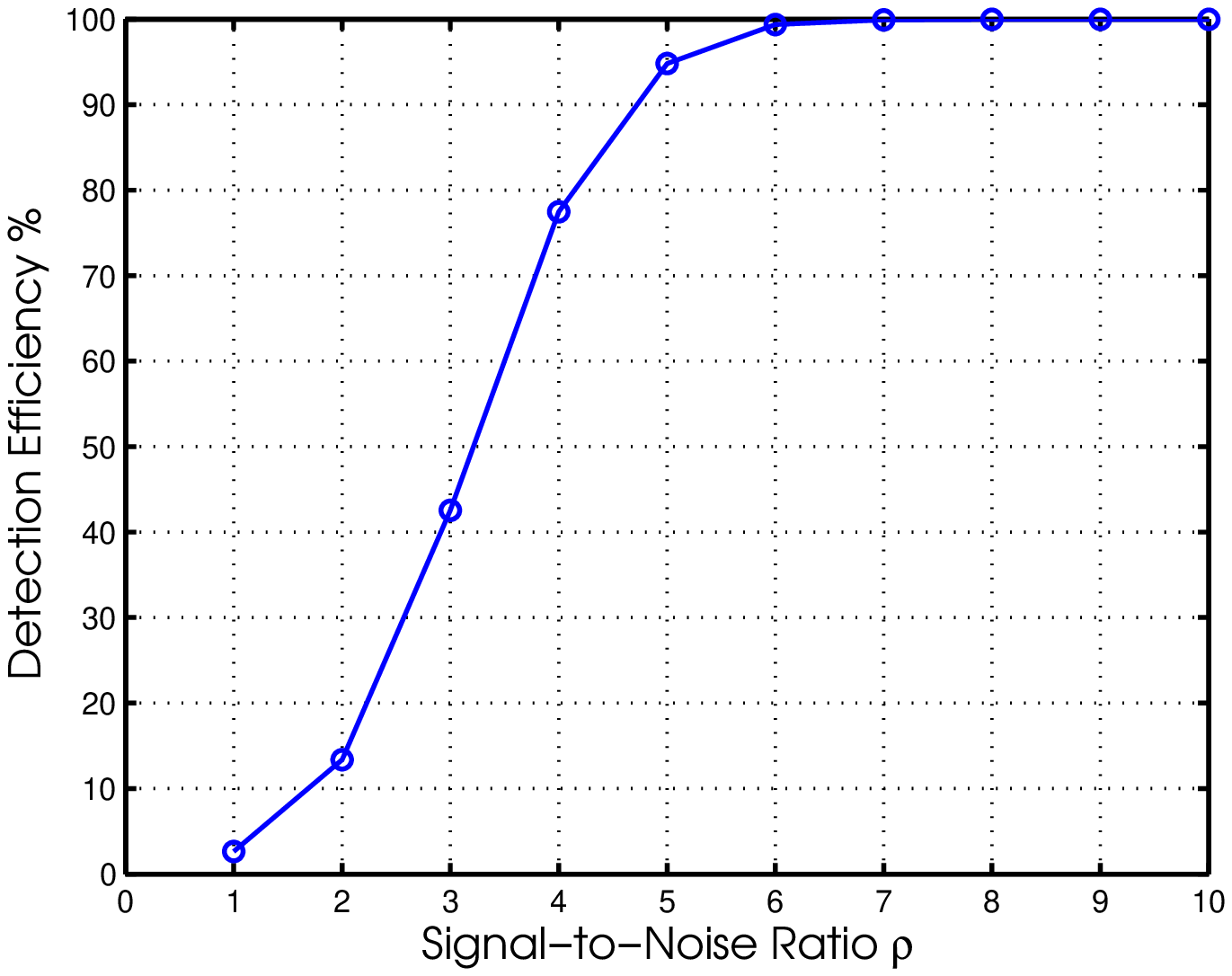}
\end{tabular}
\caption{The top figure shows the Receiver Operating Characteristic (ROC) curves
for a variety of S/N ($\rho$), with the $\rho=4$ curve showing the same form as
that given by \citet{Prix:2009}. The bottom figure shows the detection
efficiency of the known pulsar search as a function of S/N given a background
false alarm rate of 1 per cent.}
\label{fig:deteff}
\end{figure}

If we have a threshold false alarm rate due to background of 1 per cent, then
Monte Carlo simulations show that an S/N of 5 gives a detection probability of
95 per cent (see Fig.~\ref{fig:deteff}) which we will use as our required value
for confident detection in the subsequent sections. Real data is generally
non-stationary and may contain interference, so this idealised detection
statistic may be slightly optimistic, but should be close to reality, given well
understood and cleaned data.

\subsection{Second generation detectors}\label{sec:second}
\begin{figure}
\begin{tabular}{c}
\includegraphics[width=84mm]{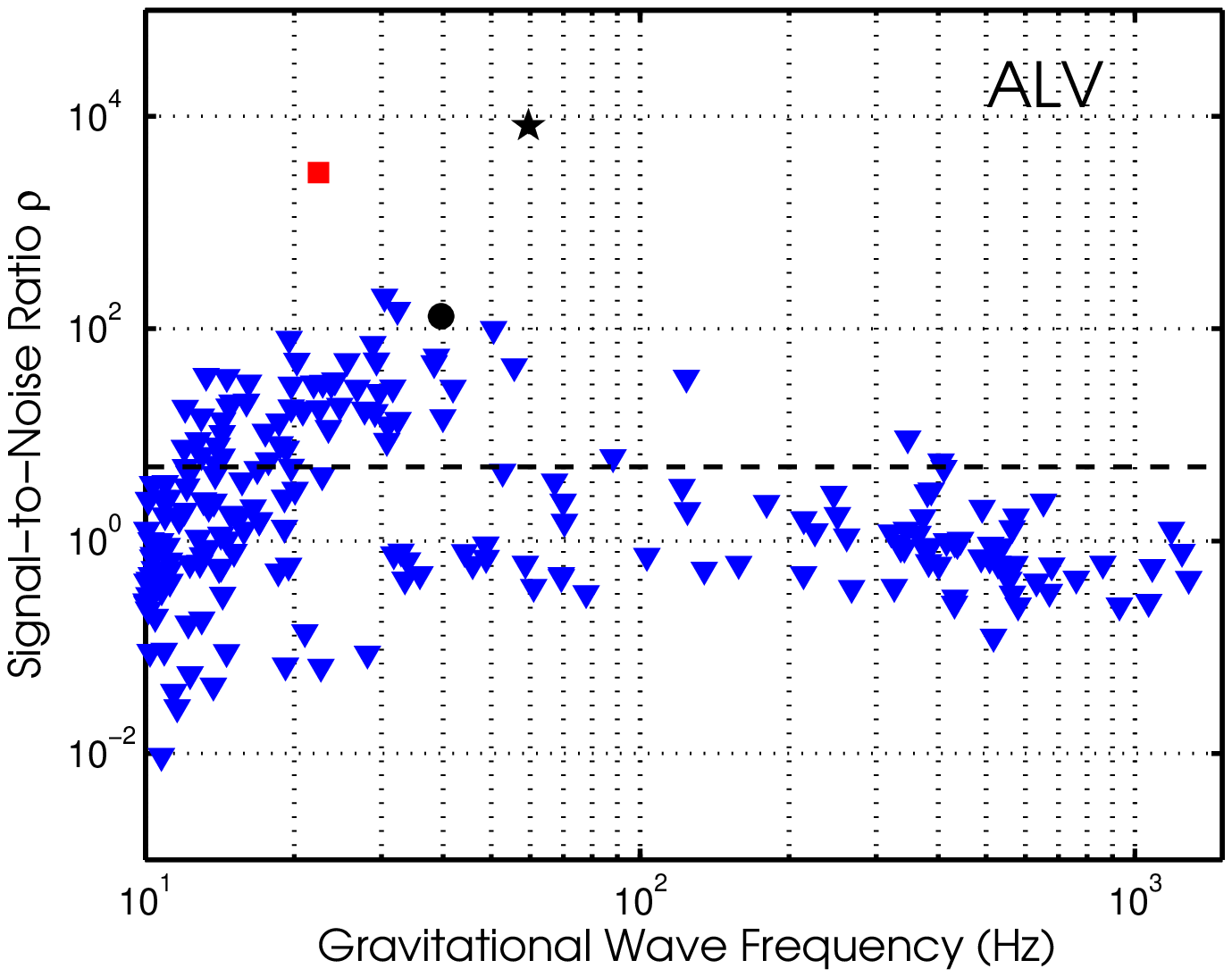} \\
\includegraphics[width=84mm]{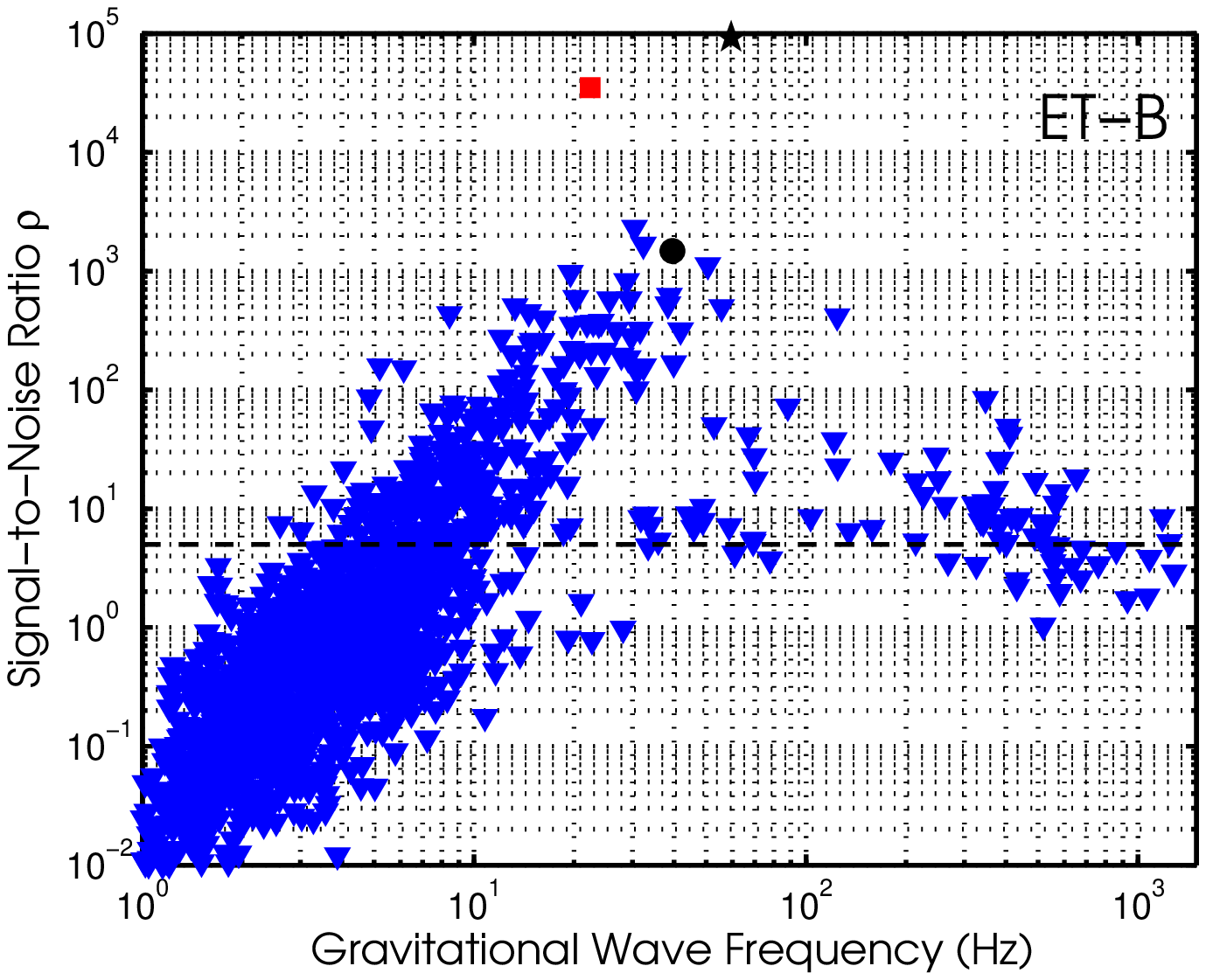} \\
\includegraphics[width=84mm]{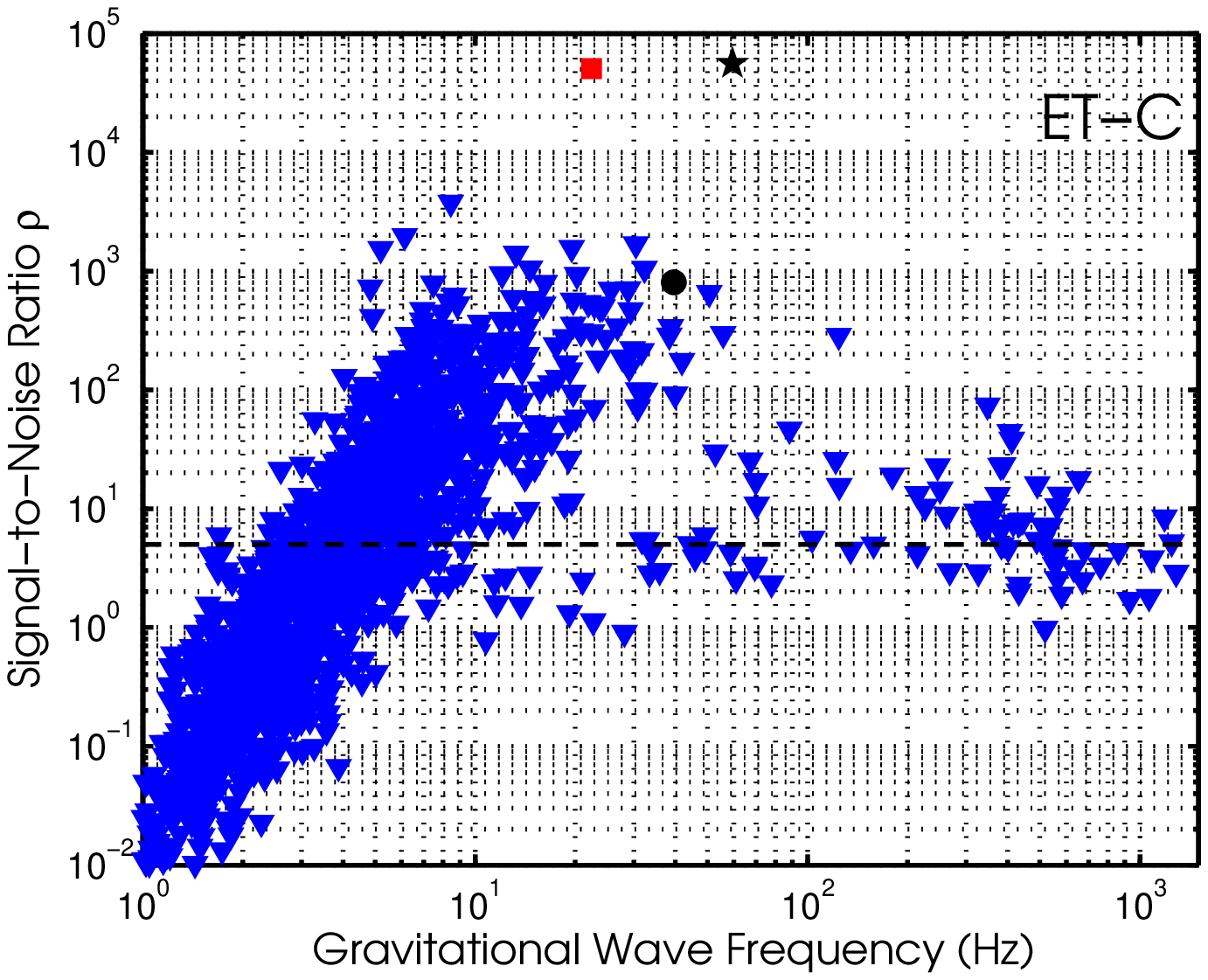}
\end{tabular}
\caption{The angle averaged signal-to-noise ratios for all currently known
non-globular cluster pulsars with measured spin-down values expected for ALV
(upper panel) and the Einstein Telescope in ET-B (middle panel) and ET-C
(bottom panel) configurations assuming emission at the spin-down limit (see text
for exceptions). The $\bigstar$ represents the value for the Crab pulsar, the
$\bullet$ represents the value for B0540$-$69, and the $\blacksquare$
represents the
value for the Vela pulsar. The thick horizontal dashed line shows an S/N of 5.}
\label{fig:AL_ET_SNRs}
\end{figure}

For aLIGO we assume two 4\,km interferometers based at the Hanford site and one
4\,km interferometer at the Livingston site all with equivalent sensitivity and
operating at the designed value \citep{LSC:2009}. For AdvVirgo we use one 3\,km
interferometer at its design sensitivity (see Fig.~\ref{fig:strains}).
Fig.~\ref{fig:AL_ET_SNRs} gives the angle averaged S/N for 1 year of
observations with all these detectors (3 aLIGO and 1 AdvVirgo, or ALV from now 
on) and shows that potentially 58 pulsars could be
observed with an S/N greater than 5 (best case orientation gives 74 with S/N
above 5 and the worst case orientation gives 47 with S/N above 5). Some similar,
but unpublished work, has been presented by \citet{Santostasi:2006}.

\subsection{Third generation detectors}\label{sec:third}
Some assessment of detectability of the currently known pulsars, based on their
spin-down limits, has previously been performed for the proposed third
generation detector called the Einstein Telescope in \citet{Andersson:2009}
(along with the potential for observing neutron stars through other \gw emission
mechanisms), although this was just for the ET-B configuration (see below).
Here, as above, we will calculate S/N for the Einstein Telescope. In
Fig.~\ref{fig:AL_ET_SNRs} we show S/N based on the sensitivity of the ET-B
configuration, as described in \citet{Hild:2008}, with three interferometers on
one site (given here as the current Virgo site) in an equilateral triangle
configuration \citep{Freise:2009}. For this configuration of ET-B we could
potentially see 312 pulsars with an angle averaged S/N greater than 5 (or 405
for the best case and 223 for worst case). A `xylophone' configuration (ET-C)
consisting of different detectors optimised for different frequency bands, with
comparable high-frequency sensitivity, but better low-frequency sensitivity than
ET-B, has also been proposed \citep{Hild:2010, Freise:2009}. In
Fig.~\ref{fig:AL_ET_SNRs} we show the S/N based on the ET-C sensitivity (again
with the same layout as ET-B). For ET-C we could potentially see 648 pulsars
with an orientation angle-averaged S/N greater than 5 (or 771 for the best case
and 528 for worst case). These numbers are around double those for the ET-B
configuration due to the lower frequency sensitivity increase, suggesting that
this may be the preferred configuration for known pulsar searches. However, we
should be cautious as we will see in \S\ref{sec:limits} (and also as stated in
\citealp{Andersson:2009}) that many of these will have to be rather exotic, and
therefore probably less realistic, stars to be seen at these levels. A further
configuration, called ET-D \citep{Hild:2010b}, is now also available, although
it is very comparable to ET-C for a wide range of frequencies, so we have not
used it in these studies.

\subsection{Parameter estimation}
Our ability to constrain the physical parameters of a source is obviously
dependent on the S/N, but the correlated physical parameters we use to describe
the signal can mean that rather unintuitively a higher S/N signal may not give
the best constraints on $h_0$.

We have simulated a selection of pulsars that give best case and worst case S/N
of around 5, 10, 50 and 100 for ALV (as shown in Fig.~\ref{fig:AL_ET_SNRs}),
to assess the level of uncertainty on estimates of their parameters assuming
they are observable at that level. To do this we used the Markov chain Monte
Carlo technique described in \citet{Abbott:2010a} to produce posterior
probability density functions (pdfs) on the four parameters defined in 
\S\ref{sec:intro}. Uncertainties on parameters can be assessed using a Fisher
Information Matrix approach, for example as done by \citet{Jaranowski:2010}, but
as they show, such a method will give {\it biased} error estimates
overestimating the uncertainty for signals near $\cos{\iota} = \pm1$, even for
very strong signals with S/N approaching 1000.

\begin{figure*}
\begin{tabular}{c c}
\includegraphics[width=84mm]{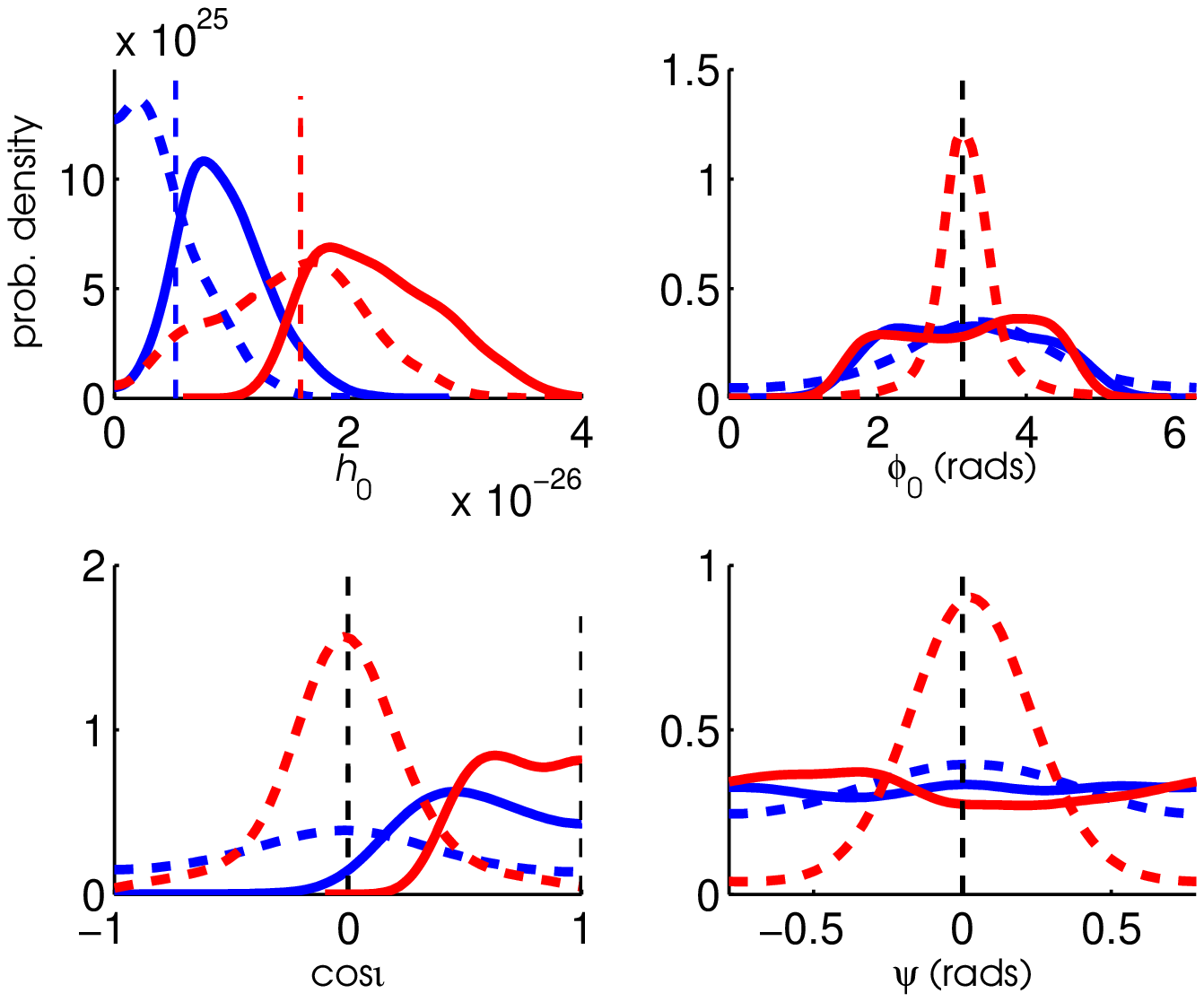} &
\includegraphics[width=84mm]{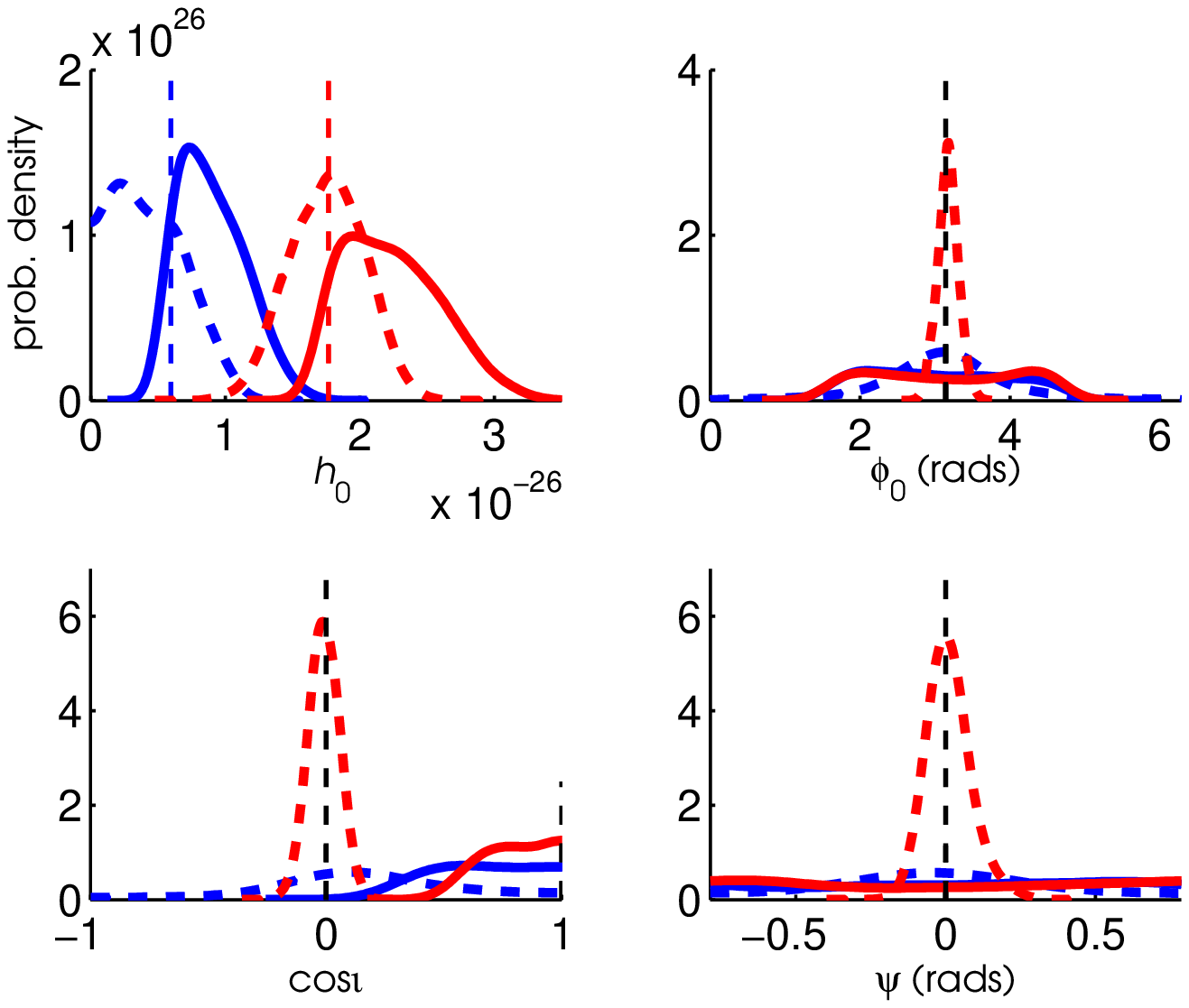} \\
\includegraphics[width=84mm]{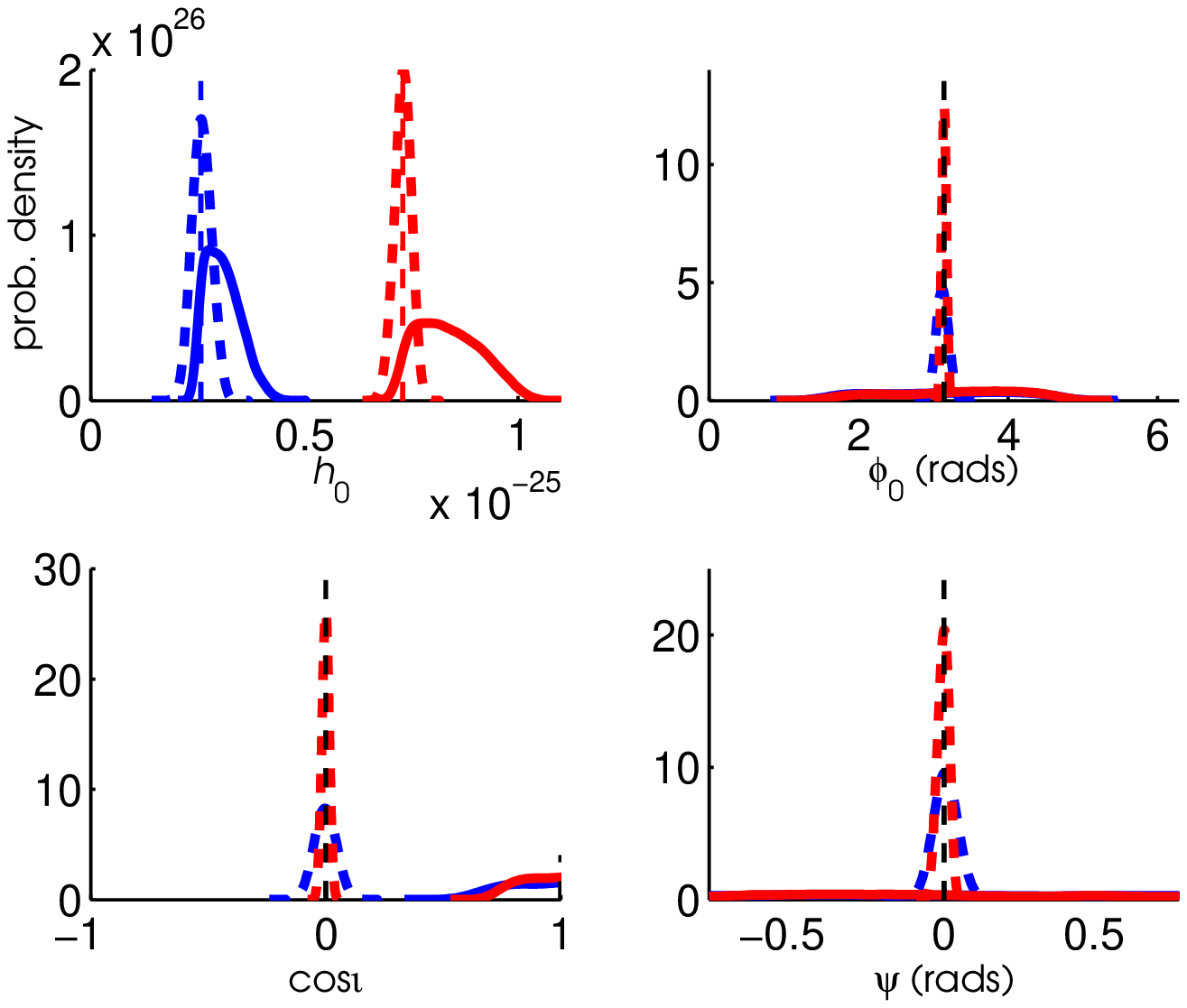} &
\includegraphics[width=84mm]{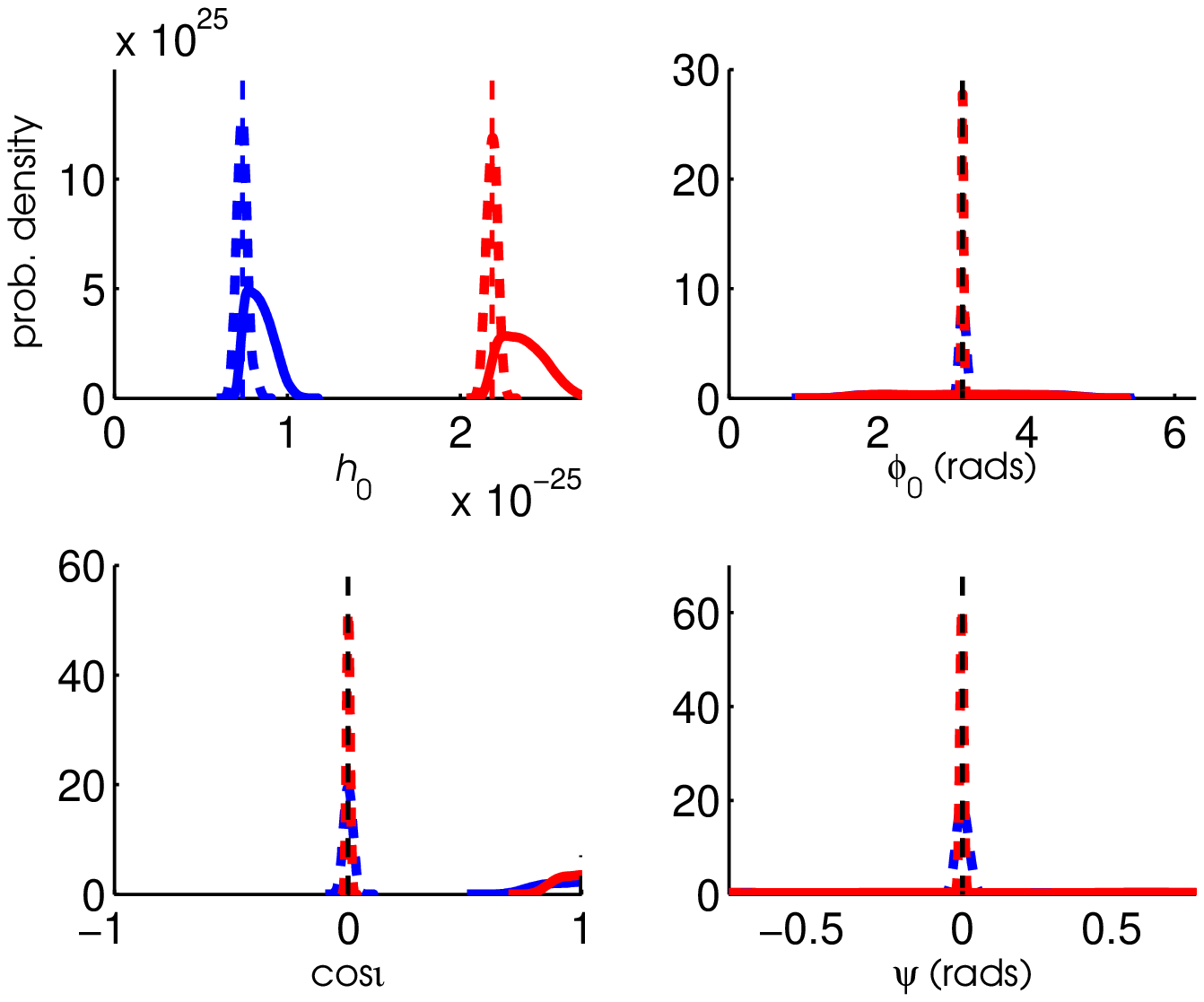}
\end{tabular}
\caption{Each panel shows the average posterior probability density functions on the four \gw 
signal parameters of $h_0$, $\phi_0$, $\cos{\iota}$ and $\psi$ for four simulated signals. 
The four signals in each panel (from top left to bottom right) correspond to the parameters 
given in each of the four sections (from top to bottom) of Table~\ref{tab:uncertainties}.
In an individual panel lines of the same colour represent signals with identical
amplitude, solid lines represent the signals with the worst case orientation 
($\cos{\iota}=0$) and dashed lines represent the signals with the best case orientation
($\cos{\iota}=1$). The tapering off of the $\phi_0$ posteriors
for the best case orientation at $\pi\pm\pi/2$ is an artifact of the plotting
and these parameters should really have abrupt cut-offs at these values.}
\label{fig:ALSNR}
\end{figure*}

We calculate the appropriate signal strength needed to give S/N of 5, 10, 50
and 100 for both the best and worst case orientations (see
Table~\ref{tab:uncertainties}) and produce pdfs on many realisations of the
data to give an average pdf. These can be seen in Fig.~\ref{fig:ALSNR} and the
mean parameter values (not the peak in the posterior) and their $1\sigma$
uncertainties obtained from the pdfs are given in Table~\ref{tab:uncertainties}.
Examples of pulsars that if emitting at their spin-down limits might be observed
with ALV at S/N of $\sim5$, 10, 50 and 100 are respectively J1959+2048,
J0737$-$3039A (the millisecond pulsar in the famous double pulsar system),
J0537$-$6910 (an interesting young pulsar in the Large Magellanic Cloud, with a
high glitch rate and large spin-down rate) and J1952+3252.
\begin{table*}
\begin{minipage}{160mm}
\caption{Mean parameter estimates, and their uncertainties, at a variety of
signal-to-noise ratios given best and worst case orientations. The true value of
$h_0$ is given followed by the average (over many trials) mean estimate and
standard deviation on each parameter. The fractional uncertainty on $h_0$ is
also given. In the low S/N cases where the signal peaks near zero a 95 per cent
confidence upper limit is given. The true values of $\phi_0$ and $\psi$ in all cases
are $\pi$\,rads and 0\,rads respectively. The parameter estimates have been
performed assuming the ALV combination at design
sensitivity.}\label{tab:uncertainties}
\begin{tabular}{c | c | c | c | c | c | c | c}
\hline
~ &  True $h_0$ & $\rho$ & $\overline{h_0} \pm \sigma_{h_0}$
& $\sigma_{h_0}/\overline{h_0}$ (per cent) & $\overline{\phi_0} \pm
\sigma_{\phi_0}$ &
$\overline{\cos{\iota}} \pm \sigma_{\cos{\iota}}$ & $\overline{\psi} \pm
\sigma_{\psi}$ \\
\hline
$\cos{\iota}$ = 0 & $5.24\ee{-27}$ & 1.7 & $h_0^{95~{\rm per~cent}} =
1.08\ee{-26}$ & * &
$3.28 \pm 1.24$ & $-0.02 \pm 0.46$ & $0.00 \pm 0.40$ \\
$\cos{\iota}$ = 1 & $5.24\ee{-27}$ & 5.0 & $9.27 \pm 3.72\ee{-27}$ & 40 &
$3.21$ & $0.51 \pm 0.26$ & * \\
$\cos{\iota}$ = 0 & $1.59\ee{-26}$ & 5.0 & $1.49 \pm 0.62\ee{-26}$ & 42 &
$3.20 \pm 0.44$ & $0.00 \pm 0.29$ & $0.03 \pm 0.21$ \\
$\cos{\iota}$ = 1 & $1.59\ee{-26}$ & 15.1 & $2.27 \pm 0.55\ee{-26}$ & 24 &
$3.19$ & $0.67 \pm 0.18$ & * \\
\hline
$\cos{\iota}$ = 0 & $5.96\ee{-27}$ & 3.4 & $h_0^{95~{\rm per~cent}} =
9.15\ee{-27}$ & * &
$3.12 \pm 0.90$ & $0.16 \pm 0.36$ & $-0.02 \pm 0.33$ \\
$\cos{\iota}$ = 1 & $5.96\ee{-27}$ & 10.0 & $8.98 \pm 2.49\ee{-27}$ & 28 &
$3.06$ & $0.64 \pm 0.21$ & * \\
$\cos{\iota}$ = 0 & $1.77\ee{-26}$ & 10.0 & $1.76 \pm 0.27\ee{-26}$ & 15 &
$3.17 \pm 0.14$ & $-0.01 \pm 0.06$ & $0.01 \pm 0.07$ \\
$\cos{\iota}$ = 1 & $1.77\ee{-26}$ & 29.6 & $2.24 \pm 0.36\ee{-26}$ & 16 &
$3.15$ & $0.78 \pm 0.14$ & * \\
\hline
$\cos{\iota}$ = 0 & $2.58\ee{-26}$ & 17.7 & $2.60 \pm 0.22\ee{-26}$ & 8 &
$3.12 \pm 0.08$ & $0.00 \pm 0.04$ & $0.01 \pm 0.04$ \\
$\cos{\iota}$ = 1 & $2.58\ee{-26}$ & 50.0 & $3.10 \pm 0.40\ee{-26}$ & 13 &
$3.19$ & $0.82 \pm 0.11$ & * \\
$\cos{\iota}$ = 0 & $7.31\ee{-26}$ & 50.0 & $7.30 \pm 0.19\ee{-26}$ & 3 &
$3.14 \pm 0.03$ & $0.00 \pm 0.01$ & $0.00 \pm 0.02$ \\
$\cos{\iota}$ = 1 & $7.31\ee{-26}$ & 141.5 & $8.40 \pm 0.73\ee{-26}$ & 9 &
$3.26$ & $0.87 \pm 0.08$ & * \\
\hline
$\cos{\iota}$ = 0 & $7.40\ee{-26}$ & 33.8 & $7.39 \pm 0.33\ee{-26}$ & 4 &
$3.15 \pm 0.04$ & $0.00 \pm 0.02$ & $0.00 \pm 0.02$ \\
$\cos{\iota}$ = 1 & $7.40\ee{-26}$ & 100.0 & $8.41 \pm 0.74\ee{-26}$ & 9 &
$3.14$ & $0.88 \pm 0.08$ & * \\
$\cos{\iota}$ = 0 & $2.18\ee{-25}$ & 100.0 & $2.18 \pm 0.03\ee{-25}$ & 1 &
$3.14 \pm 0.01$ & $0.00 \pm 0.01$ & $0.00 \pm 0.01$ \\
$\cos{\iota}$ = 1 & $2.18\ee{-25}$ & 294.8 & $2.37 \pm 0.12\ee{-25}$ & 5 &
$3.12$ & $0.92 \pm 0.05$ & * \\
\hline
\end{tabular}
\end{minipage}
\end{table*}
The main results of Table~\ref{tab:uncertainties}, i.e.\ the relative
uncertainties in $h_0$ for different S/N signals given the best and worst case
orientation scenarios, are extrapolated to higher S/N and shown as the thick
black lines in Fig.~\ref{fig:errorvssnr} (they are equivalent to the
uncertainties on the quadrupole assuming the distance is precisely known).

\begin{figure}
\includegraphics[width=84mm]{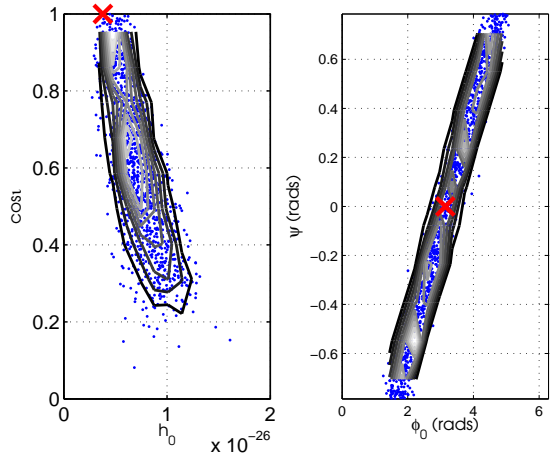}
\caption{The correlations between the $h_0$ and $\cos{\iota}$, and $\phi_0$ and
$\psi$ parameters for a simulated signal with the best case orientation. The
data points from the Markov chain Monte-Carlo (MCMC) and the probability
contours derived from them are shown. The cross marks the parameters used to
produce the signal.}
\label{fig:correlations}
\end{figure}

It can clearly be seen by looking at Eqn.~\ref{eq:signal}, and from our
posteriors in Fig.~\ref{fig:ALSNR}, that for a signal with $\cos{\iota} \approx
\pm1$ the $h_0$ and $\cos{\iota}$ parameter are highly correlated, which leads
to the $h_0$ posterior being spread to higher values, and $\cos{\iota}$
extending towards zero. In such a case the $\phi_0$ and $\psi$ parameters are
also very highly correlated, with the posterior on $\psi$ being flat over its
range (see Fig.~\ref{fig:ALSNR}), and $\phi_0$ only constrainable to within a
$180^{\circ}$ range i.e.\ the direction the star is rotating in can
be found, but otherwise these parameters are ill defined to represent such a
situation. These correlations can be seen more easily in
Fig.~\ref{fig:correlations}, and they lead to the slightly paradoxical fact that
when comparing equivalent signals in Fig.~\ref{fig:ALSNR} we can actually place
the tightest constraints on $h_0$ for the worst case orientation, i.e.\ lower
S/N, signals! Due to these correlations we do not quote mean, or uncertainty,
values on $\psi$ in Table~\ref{tab:uncertainties} as they are meaningless, we
also only quote mean values for $\phi_0$ to show that the value found is
oriented in the correct sense. If we were to cast the the uncertainties in
$\iota$ and $\psi$ as a solid angle then at $\cos{\iota} = \pm 1$ the associated
error ellipses for different S/N would be entirely proportional to the error on
$\iota$ alone (i.e\ the error on psi stays constant). For a signal with S/N
$\approx 20$ the best and worst case solid angles covering the $\sim 1\sigma$
probability contour in $\iota-\psi$ space are 0.35\,sr and $6\ee{-4}$\,sr
respectively. The ratio of these areas scales very roughly linearly with S/N
(within about a factor of 2) as $50\rho$, up to S/N of $\sim 100$.

\begin{figure}
\includegraphics[width=84mm]{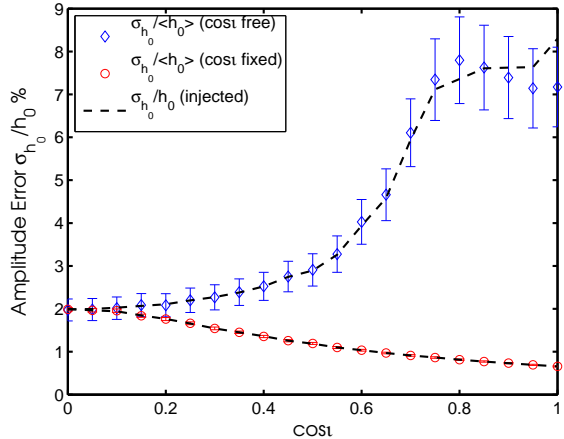}
\caption{The fractional uncertainties on amplitude (standard deviation
over mean - $\sigma_{h_0}/\overline{h_0}$) from many simulations of
performing parameter estimation on a signal (with an S/N of 50 calculated for
the worst case orientation) when all parameters are unknown (blue) compared to
when the $\cos{\iota}$ and $\psi$ parameters are known and fixed in the search.
Also shown as dashed lines are the fractional uncertainties given the actual
signal $h_0$ values rather than the mean recovered value.}
\label{fig:free_vs_fixed}
\end{figure}

In the case of pulsars like the Crab pulsar where X-ray observations and
modelling of their pulsar wind nebulae are possible, there is additional
information placing tight constraints on the orientation and polarisation
angles. Including this will break the degeneracies between parameters, for
example as perform in the search by \citet{Abbott:2008a}. In
Fig.~\ref{fig:free_vs_fixed} the fractional uncertainty on the amplitude
estimate (averaged over many simulations) is shown as a function of
$\cos{\iota}$ for simulated sources where in one case the orientation and
polarisation angles are known (i.e.\ these are fixed at their known values and
not searched over, equivalent to having very tight prior distributions on these
parameters), and in the other case that they are unknown (i.e.\ the whole range
of $\cos{\iota}$ and $\psi$ has to be searched over with uniform priors).
It can be seen in Fig.~\ref{fig:free_vs_fixed} that for the case where all
parameters are searched over the fractional uncertainty on the amplitude
increases with $\cos{\iota}$, with the steepest rise between about
$\cos{\iota}=0.5$--0.7. However, when $\cos{\iota} \gtrsim 0.8$ the fractional
error slightly decreases again suggesting that a $\cos{\iota}=\pm1$
does not actually given the worst estimate of $h_0$. This is actually slightly
misleading as the reason for the downturn becomes more apparent from
looking at the curves of fractional uncertainty given the true $h_0$ values
(rather than the fractional uncertainty based on the mean $h_0$ value), and
plots of pdfs for $h_0$ in Fig.~\ref{fig:ALSNR} -- at values of $\cos{\iota}$
approaching $\pm1$ the mean estimate of $h_0$ starts to increase relative to
the true value, whilst the uncertainty starts to plateau, leading to the
downturn. This means that estimates of $h_0$ when $\cos{\iota}=\pm1$ would
still give values furthest from the true value, but due to the shifting of the
whole posterior towards higher values rather than it widening. For the case
where the $\cos{\iota}$ and $\psi$ values are held fixed at their known values
during the parameter estimation the fractional amplitude uncertainty behaves as
one would expect and decreases as the S/N increases as $\cos{\iota}$ tends to 1.
The fractional amplitude uncertainty, whether calculated with the mean or true
$h_0$ value follow the same path, because the posterior stays symmetrical about
the true value.

Current detectors have systematic uncertainties in amplitude and phase due to
calibration in the region of 10--15 per cent and $5^{\circ}$ respectively
\citep{Abbott:2010b}. For second generation detectors it is hoped to further
reduce amplitude uncertainty to less than 10 per cent, but if we are lucky
enough to
see any high S/N ($\gtrsim 20$) signals this may be the main source of error 
in the $h_0$ estimates.

\subsubsection{Dependence on distance}
The \gw amplitude, and its uncertainties, can be directly measured. However, if
we want to convert this into a physical quantity related to the star, such as
the mass quadrupole moment (see Eqn.~\ref{eq:qmom}) then the uncertainty on the
distance to the pulsar will come into play. Current distance measurements for
nearby pulsars come from parallax measurements, but for most others come from
dispersion measure observations extrapolated from a model of the Galactic free
electron distribution \citealt{Cordes:2002}. For the majority of pulsars
distances generally have uncertainties at $\sim2$ per cent or more (even $>
100$ per cent) of the best fit values. In the future, with instruments like the
SKA, direct parallax measurements may push these errors down to the level of
$<20$ per cent for many millisecond pulsars out to 9\,kpc \citep{Smits:2010}.
Here we will look at how different distance uncertainties affect estimates of
the quadrupole moment for a variety of S/N and for the best and worst
orientations.

\begin{table*}
\begin{minipage}{130mm}
\caption{The uncertainties on the quadrupole moment at a variety of S/N given
best and worst case orientations with uncertainties on the distance measurement
of $\sigma_r =$ 1, 5, 10, 20 and 50 per cent.}\label{tab:disterr}
\begin{tabular}{c | c | c | c | c | c | c | c}
\hline
~ & ~ & ~ & \multicolumn{5}{c}{$\sigma_{Q_{22}}/\overline{Q_{22}}$ (per cent)}
\\
\hline
~ & $\rho$ & $\sigma_{h_0}/\overline{h_0}$ (per cent) or $\sigma_r = 0$ per cent
& $\sigma_r= 1$ per cent & $\sigma_r = 5$ per cent & $\sigma_r = 10$ per cent &
$\sigma_r = 20$ per cent & $\sigma_r = 50$ per cent \\
\hline
$\cos{\iota}$ = 0 & 1.7 & * & * & * & * & * & * \\
$\cos{\iota}$ = 1 & 5.0 & 40 & 41 & 40 & 43 & 46 & 59 \\
$\cos{\iota}$ = 0 & 5.0 & 42 & 42 & 44 & 45 & 49 & 60 \\
$\cos{\iota}$ = 1 & 15.1 & 24 & 25 & 25 & 27 & 32 & 44 \\
\hline
$\cos{\iota}$ = 0 & 3.4 & * & * & * & * & * & * \\
$\cos{\iota}$ = 1 & 10.0 & 28 & 29 & 29 & 30 & 35 & 50 \\
$\cos{\iota}$ = 0 & 10.0 & 15 & 16 & 17 & 19 & 25 & 38 \\
$\cos{\iota}$ = 1 & 29.6 & 16 & 17 & 17 & 20 & 26 & 41 \\
\hline
$\cos{\iota}$ = 0 & 17.7 & 8 & 9 & 10 & 13 & 21 & 40 \\
$\cos{\iota}$ = 1 & 50.0 & 13 & 13 & 14 & 17 & 23 & 40 \\
$\cos{\iota}$ = 0 & 50.0 & 3 & 3 & 6 & 11 & 19 & 36 \\
$\cos{\iota}$ = 1 & 141.5 & 9 & 9 & 10 & 13 & 21 & 38 \\
\hline
$\cos{\iota}$ = 0 &  33.8 & 4 & 5 & 7 & 11 & 19 & 36 \\
$\cos{\iota}$ = 1 & 100.0 & 9 & 9 & 10 & 13 & 20 & 39 \\
$\cos{\iota}$ = 0 & 100.0 & 1 & 2 & 5 & 10 & 18 & 36 \\
$\cos{\iota}$ = 1 & 294.8 & 5 & 6 & 7 & 11 & 20 & 39 \\
\hline
\end{tabular}
\end{minipage}
\end{table*}

\begin{figure}
\begin{tabular}{c}
\includegraphics[width=84mm]{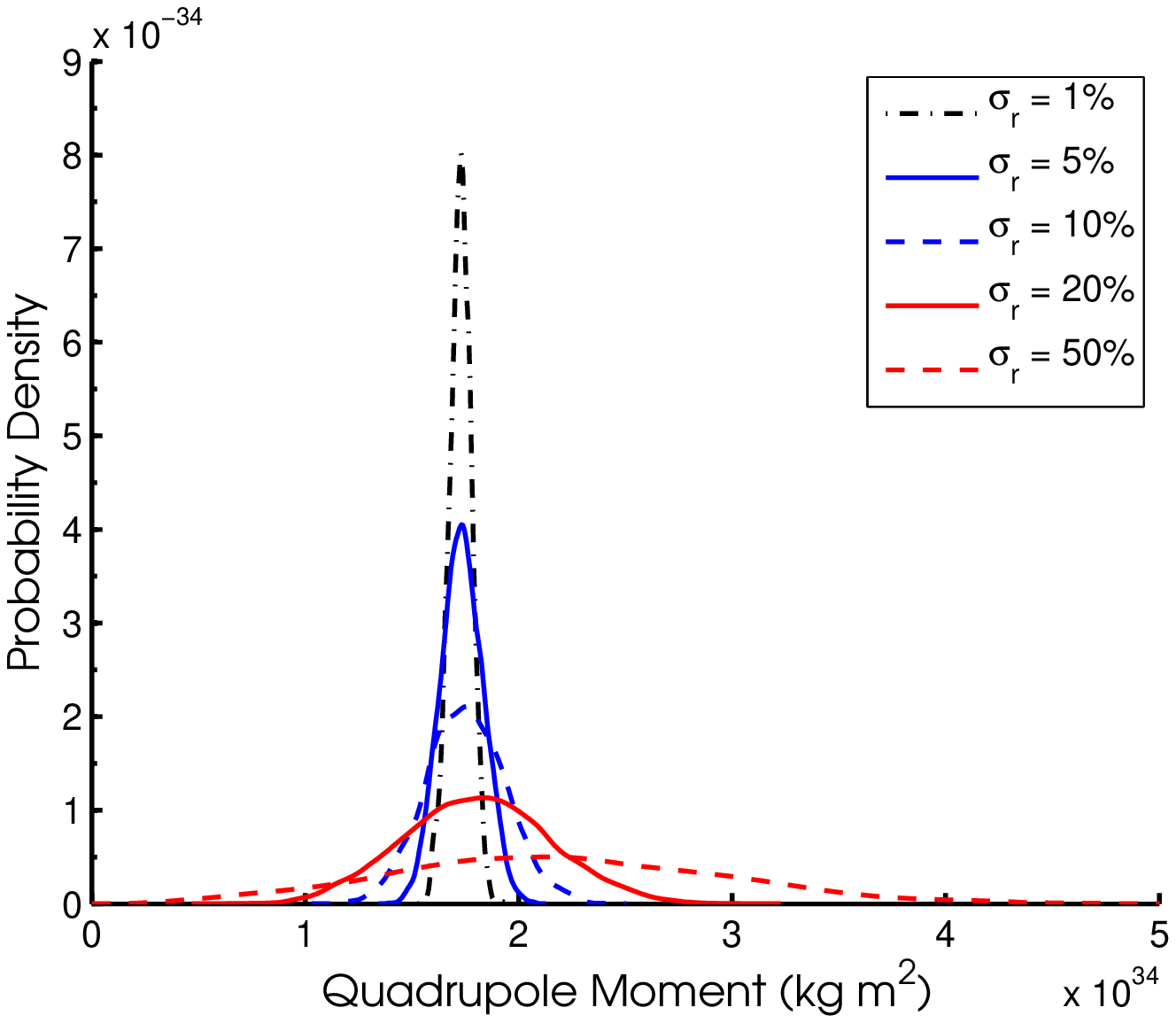} \\
\includegraphics[width=84mm]{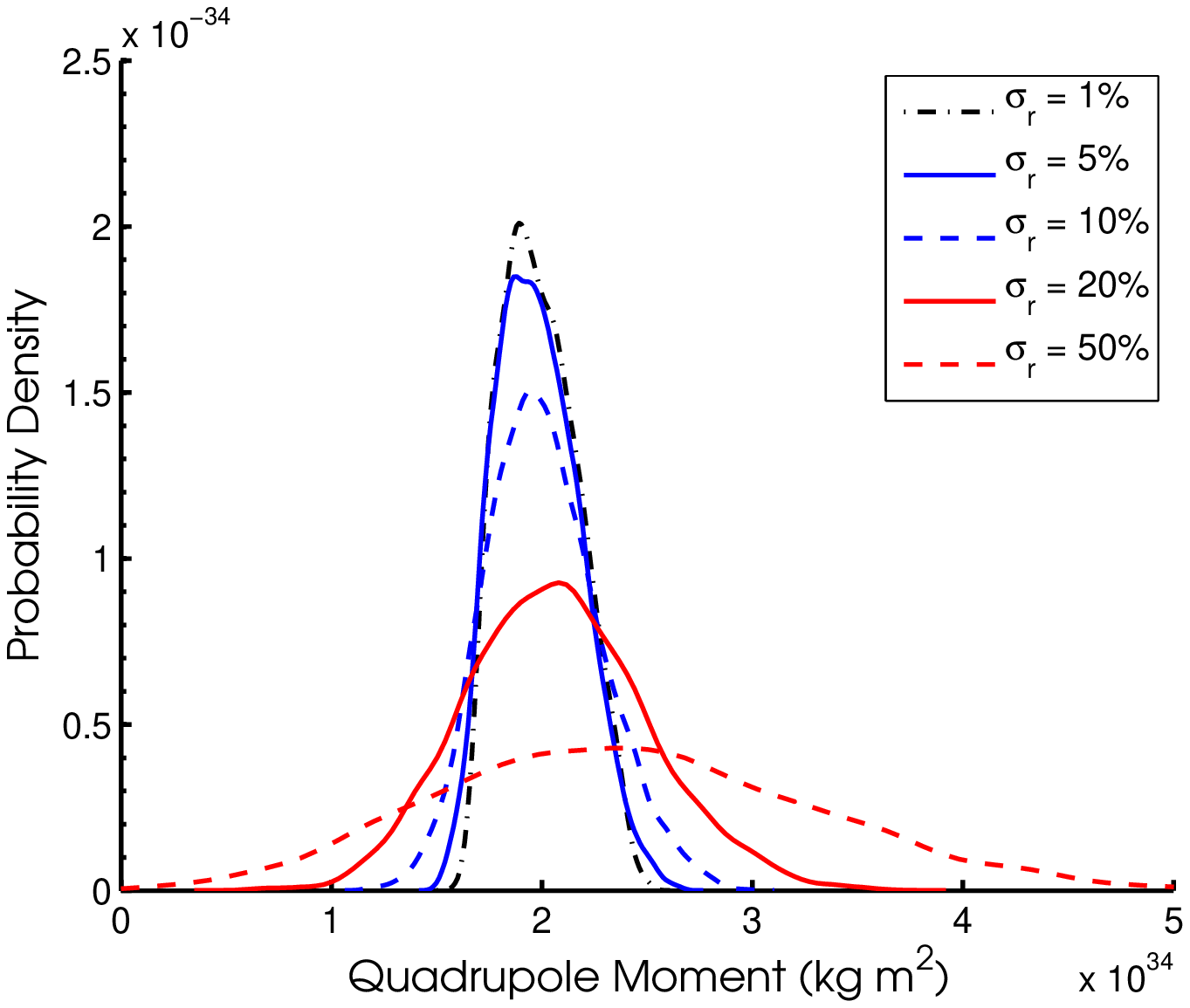}
\end{tabular}
\caption{The pdf for the quadrupole moment, $Q_{22}$, given a simulated
signal for the pulsar with $h_0 = 7.31\ee{-26}$ in
Table~\ref{tab:uncertainties} at both $\cos{\iota} = 0$ (top plot) and
$\cos{\iota} = 1$ (bottom plot), with S/N of 50 and 141.5 respectively, over a
range of distance uncertainties, $\sigma_r$, of 1, 5, 10, 20 and 50 per cent.}
\label{fig:disterr}
\end{figure}

\begin{figure}
\includegraphics[width=84mm]{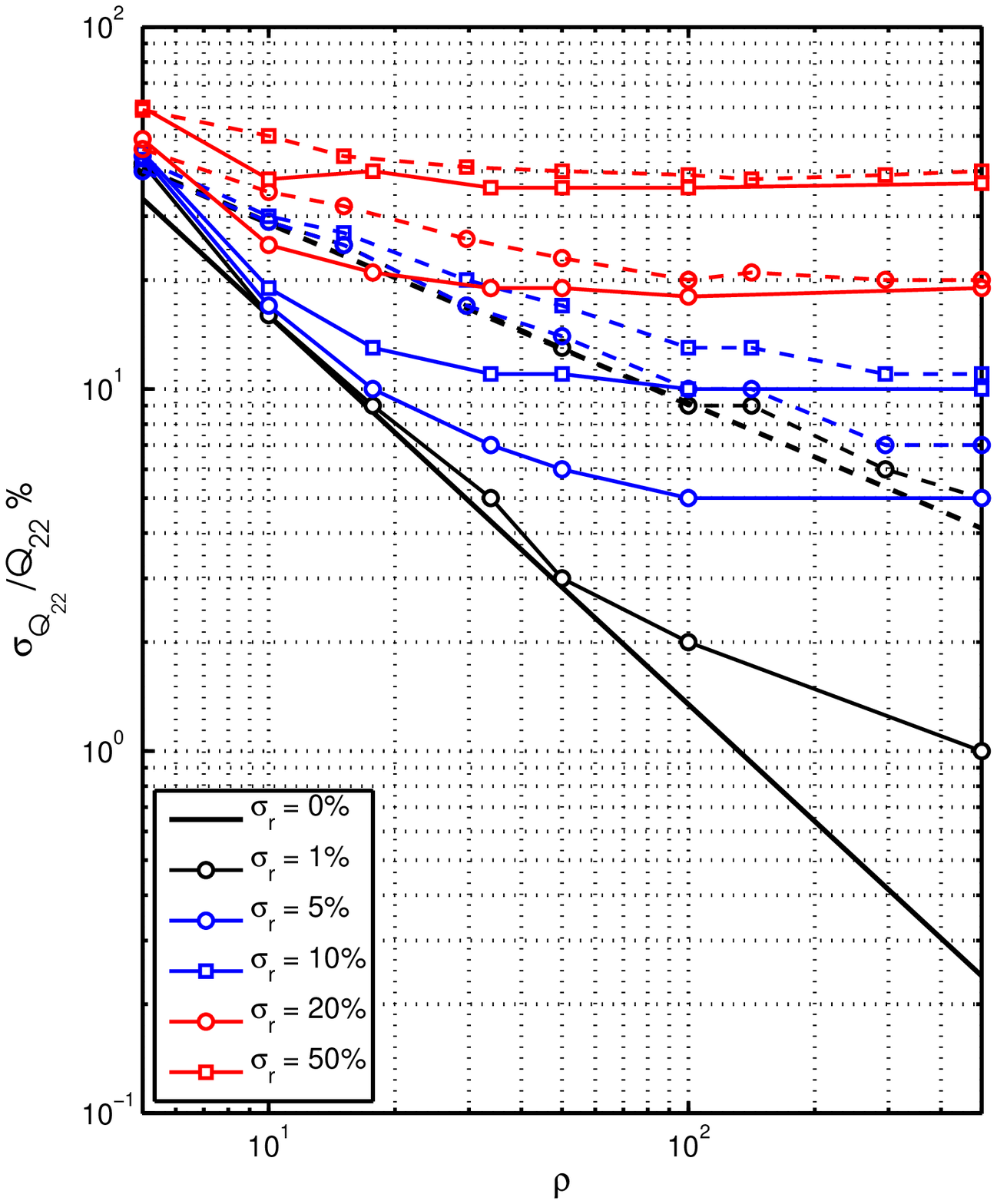}
\caption{The uncertainties on the estimate of $Q_{22}$ given in
Table~\ref{tab:disterr} for a variety of distance uncertainties and the best
(dashed lines) and worst (solid lines) case orientations.}
\label{fig:errorvssnr}
\end{figure}

It can be seen from Table~\ref{tab:disterr} and Fig.~\ref{fig:disterr} that for
the lowest S/N signals distance errors of 10--20 per cent give only a relatively
small
increase in the uncertainty with which the quadrupole moment can be measured.
The convergence of the uncertainties on the measurement for the best and worst
case orientations, i.e.\ when the uncertainties become dominated by the distance
uncertainty rather than that on the measurement, can be seen in
Fig.~\ref{fig:errorvssnr}. For strong signals (S/N $\gtrsim 50$) the uncertainty
on distance will start to dominate quadrupole estimates if it is $\gtrsim 10$
per cent and will also swamp any differences due to orientation. 

\citet{Seto:2005} suggests that \gw observations of pulsars could be used to
determine their distances to $\lesssim10$ per cent, but only for nearby stars
($\lesssim 100$ pc) with very large S/N and high frequencies. This is therefore
unlikely to help constrain distances better than other techniques for the vast
majority of known pulsars.


\section{Equations of state}\label{sec:eos}
As discussed in \citet{Lattimer:2007} there are many ways to attempt to
observationally constrain the \eos of a neutron star through inferences about
their mass and radius e.g.\ via various observations of binary and accreting
systems. Observations of \gws from merging neutron stars (or merging
neutron stars and black holes) can also give mass and radius measurements, and
information about the \eos properties from tidal effects during the inspiral
\citep{Flanagan:2008} and the point at which the star breaks up
\citep{Andersson:2009}.

However, emission of \gws from individual neutron stars, either through a
continuous emission from a sustained triaxiality, or short bursts from
vibrational modes (e.g.\ \citealp{Andersson:1998}), could also provide
constraints. Due to the low amplitude of \gws expected from these sources, such
observations would most likely only be available for Galactic sources. Rather
than studying explicitly what observations of continuous \gws from a triaxial
neutron star can tell us about the stars \eos, we flip the question
and look at the maximum magnitudes of signals we could possibly expect for a
variety of \eoss. The maximum values we discuss are likely to be at
the extreme end of possibility and in reality they could overestimated (or
possibly, but less likely, underestimated) by several orders of magnitude due to
the uncertain physics of these objects.

\subsection{Quadrupole moments}
The observable in known pulsar searches that relates to the neutron
star \eos is the \gw amplitude $h_0$. This is directly related to the star's
$l=m=2$ mass quadrupole moment via
\begin{equation}\label{eq:qmom}
Q_{22} = h_0 \left(\frac{c^4r}{16\pi^2G\nu^2}\right)\sqrt{15/8\pi},
\end{equation}
where $r$ is the distance in m and $\nu$ is the rotation frequency in Hz. The
quadrupole moment is often parametrized in terms of the star's principal moment
of inertia, $I_{zz}$, and ellipticity, $\varepsilon$, via $\varepsilon I_{zz} =
\sqrt{8\pi/15}Q_{22}$ \citep{Owen:2005}. However, these two parameters cannot be
disentangled from observations, although the range of moments of inertia can be
implied from various theoretical \eos mass and radius relations
(generally thought to be in the range $1-3\ee{38}$\,kg\,m$^2$
e.g.\ \citealp{Abbott:2007a}), allowing ellipticities to be implied from
observations. Often in \gw literature results are cast in terms of ellipticity
as it feels a more physically tangible parameter, giving a physical ``size'' of
the distortion of a star. 

Following for example \citet{Ushomirsky:2000} and \citet{Owen:2005} the
quadrupole for an incompressible star with a thin crust can be written as
\begin{equation}\label{eq:quadmom}
Q_{22} = \frac{\gamma \mu R^{6} \langle\sigma_{22}\rangle}{GM},
\end{equation}
where $\langle\sigma_{22}\rangle$ is the weighted average strain on the crust 
(contributing to the 22-quadrupole), $R$ and $M$ are the star's radius and mass,
and $\gamma \approx 13$. Often this equation is quoted as an upper limit on
$Q_{22}$ by inserting the maximum breaking strain $\sigma_{\rm max}$,
however as noted in \citet{Ushomirsky:2000} the equation is equally applicable
to providing an estimate of the crustal strain. The parameter in
Eqn.~\ref{eq:quadmom} that is most dependent on the detailed make-up of the star
is the shear modulus $\mu$, which between theories can potentially vary by many
orders of magnitude. Below we will make the assumption that the stars are
incompressible and therefore use Eqn.~\ref{eq:quadmom} in calculations,
along with an assumed $\gamma=13$ (for a conventional neutron
star $\gamma$ can be related to the ratio of the thickness of the star's
crust, $\Delta{}R$, to it's radius via $\gamma\approx120\Delta{}R/R$). It
should also be noted that these assume the Cowling approximation and neglect
self-gravity of the density perturbations. \citet{Ushomirsky:2000} showed that
including self-gravity can increase the quadrupole calculations by between
25--200 per cent, and similarly \citet{Haskell:2006} found that self gravity can
affect
results by factors of between 0.5--3.

For the millisecond recycled pulsars spin-down arguments alone tell us that
their quadrupoles must be relatively small ($\lesssim 10^{30}$kg\,m$^2$). Such
a quadrupole would be obtainable with any \eos (see below) meaning
that if detected they are not helpful differentiating between theories, although
multiple detections could build up useful statistics on their properties and
limits on their internal magnetic fields.

Here we will review some of the work presented by \citet{Owen:2005} regarding
maximum sustainable quadrupoles for a variety of stellar \eoss.

\subsubsection{Normal neutron stars}\label{sec:normalns}
For stars made from {\it normal} neutron star matter (neutrons, protons and
electrons) \citet{Ushomirsky:2000} provide a detailed model of the quadrupole
(see Eqn. 69). \citet{Owen:2005} applies standard numbers in this definition
(and corrects the definition of the shear modulus to be $4\ee{29}\,{\rm
erg}\,{\rm cm}^{-3}$, or $2.5\ee{-4}$\,MeV\,fm$^{-3}$) to give
\begin{equation}\label{eq:quadmom2}
Q_{22} = 2.4\ee{32}\,{\rm kg}\,{\rm m}^2 \langle\sigma_{0.1}\rangle
R_{10}^{6.26} M_{1.4}^{-1.2},
\end{equation}
where $\langle\sigma_{0.1}\rangle$ is an averaged strain of $0.1$, $R_{10}$ is
the radius in units of 10\,km, and $M_{1.4}$ is the mass in units of
$1.4\,M_{\odot}$. Uncertainties in the star's mass will only affect this
estimate by small amounts e.g.\ given theoretical\footnote{Theoretically the
lower bound on neutron star mass could be as small as $\sim 0.1\,M_{\odot}$
\citep{Lattimer:2001}, but we will assume our population of known pulsars is
similar to the ones with observed masses -- see e.g.\ Fig. 3 of
\citet{Lattimer:2007}, or the figure maintained at 
\url{http://www.stellarcollapse.org/nsmasses}. A discussion of \gws from
neutron stars at the lowest end of the possible mass range can be found in 
\citet{Horowitz:2009b}, which suggest these are potentially strong sources, but 
it is very unlikely that any known pulsar would be of this type.} and 
observational bounds on the mass between $\sim 1-2.5\,M_{\odot}$ the quadrupole
will only vary within about a factor of three from 0.5--1.5. However, for the
radius, with its far larger exponent, small differences can give a larger range
of possible quadrupoles. If we take a theoretical range from 10--15\,km then
this can change the quadrupole by about an order of magnitude. The most massive
stars will also have the smallest radii, so from this we get an uncertainty
range on the quadrupole from the unknown mass and radius of between
$\sim0.5$--20 times the value in Eqn.~\ref{eq:quadmom2}, with the most massive,
but smallest stars at the lower end and vice versa. 

Here we will assume the breaking strain is at the maximum value of $\sigma_{\rm
max} \approx 0.1$ calculated by \citet{Horowitz:2009} (much previous work has
assumed a maximum breaking strain of $10^{-5} \le \sigma_{\rm max} \le
10^{-2}$). This value of the breaking strain was calculated for normal neutron
star matter, but for other situations it may well not be a valid assumption.
Using the higher value from the mass/radius uncertainty as an upper limit, and
inserting in the maximum breaking strain, we {\it could} get normal neutron
stars with quadrupoles of $Q_{\rm max} \approx 4.5\ee{33}\,{\rm kg}\,{\rm m}^2$.
Converting this to an approximate (order of magnitude) estimate of the
ellipticity, assuming the canonical moment of inertia, would give $\varepsilon
\approx 6\ee{-5}$. Although the reasoning behind it is quite different (i.e.\
just coming from plugging in masses and radii at the extent of their ranges,
with the majority of the increase over the fiducial value in \citet{Owen:2005}
coming from using a maximum radius of 15\,km) this value is very similar to that
produced by the perturbative approach to the problem performed by
\citet{Haskell:2006}. Assuming the maximum breaking strain of 0.1 they would
produce a maximum quadrupole (see Table~4 of \citealp{Haskell:2006}) of $Q_{\rm
max} = 3.1\ee{33}\,{\rm kg}\,{\rm m}^2$ for a star with a mass of
1.4\,$M_{\odot}$, radius 12.3\,km and crust thickness 1.5\,km.

\subsubsection{Hybrid crystalline colour-superconducting star}\label{sec:css}
In \citet{Knippel:2009}, \citet{Haskell:2007} and \citet{Lin:2007} the
quadrupole is calculated for crystalline colour-superconducting (CSS) hybrid
stars. In these stars the \gw emission mainly comes from a deformed interior
core of quark matter. The quadrupole can again be approximated by
Eqn.~\ref{eq:quadmom}, but with a shear modulus given by \citep{Mannarelli:2007}
\begin{equation}
\mu = 2.47\,{\rm MeV}\,{\rm fm}^{-3}\Delta_{10}^2\mu_{q~400}^2,
\end{equation}
where $\Delta_{10}$ is the gap parameter in units of 10\,MeV, and $\mu_{q~400}$
is the quark chemical potential in units of 400\,MeV (which in
\citealp{Mannarelli:2007} is estimated to be in the range $350\,{\rm MeV} \le
\mu_q \le 500\,{\rm MeV}$), and the stellar mass and radius are replaced by
those of the quark core. \citet{Knippel:2009} assume a range of gap
parameters\footnote{\citet{Mannarelli:2007} gives a range from $5\le
\Delta \le 25$\,MeV, although \citet{Knippel:2009} suggest that for the low
temperature CCS phase appropriate for these cores larger gap values, maybe up
to 100\,MeV, are possible.} $10\le \Delta \le 50$\,MeV, and find a maximum
core mass of 0.8\,$M_{\odot}$ and a maximum core radius of 7\,km. For the
reasons stated in \S\ref{sec:normalns} we will assume a slightly more
conservative maximum breaking strain than that for normal neutron stars of
$\sigma_{\rm max}=10^{-2}$. This gives a maximum quadrupole (for
$\Delta = 50$\,MeV, $\mu_q = 500$, $\mu = 97$\,MeV\,fm$^{-3} \approx
1.5\ee{34}\,{\rm J}\,{\rm m}^{-3}$) of $Q_{\rm max} \approx 1.4\ee{36}\,{\rm
kg}\,{\rm m}^2$, or almost three orders of magnitude larger than a {\it normal}
neutron star, which is not necessarily surprising since deformations are in the
high density core rather than the crust. \citet{Haskell:2007} note that for a
star with a fluid envelope around the core the quadrupole will be suppressed,
particularly if there is not a substantial change in density when transitioning
between the core and envelope. Converting this to an approximate estimate of the
ellipticity, by assuming the canonical moment of inertia, would give an
equivalently large $\varepsilon \approx 0.02$!

\subsubsection{Hybrid and meson condensate stars}
\citet{Owen:2005} also looks at hybrid stars with charged meson condensates and
quark-baryon cores. For these the shear modulus is given by
\begin{equation}
\mu = 0.25\,{\rm MeV}\,{\rm fm}^{-3} q_{0.4}^2 D_{15}^6 S_{30}^{-4},
\end{equation}
where $q_{0.4}$ is the charge density of quark droplets in units of -0.4$e$,
$D_{15}$ is their diameter in units of 15\,fm, and $S_{30}$ is their spacing in
units of 30\,fm. Following the correction for charge screening
in \citet{Owen:2005} an upper limit on shear modulus is given as $\mu \approx
1.3\ee{-2}\,{\rm MeV}\,{\rm fm}^{-3}$. If we evaluate Eqn.~\ref{eq:quadmom}
with this, again substituting for the core radius (which as in \S\ref{sec:css}
we will set as 8\,km) and using a fiducial stellar mass of 1.4\,$M_{\odot}$ and
a maximum breaking strain of $\sigma_{\rm max} = 10^{-2}$, we get an upper limit
on the quadrupole moment of $Q_{\rm max} \approx 3.5\ee{32}\,{\rm kg}\,{\rm
m}^2$. This is less than the extremal value for a {\it normal} neutron star. The
work by \citet{Kurkela:2010a, Kurkela:2010b} into cold quark matter suggest that
hybrid stars (with pure phases of hadronic and quark matter) could
have masses up to $\sim 2.1\,M_{\odot}$ and radii of $\sim 13$\,km although
these are likely to have only a tiny quark core i.e.\ they will mainly look like
a normal hadronic neutron star. However, they show that stars with mixed phases
of quarks and hadrons can have masses of up to $\sim 1.9\,M_{\odot}$ and radii
of $\sim 11$\,km, which with the above assumptions allows larger quadrupole
moments of $Q_{\rm max} \approx 1.8\ee{33}\,{\rm kg}\,{\rm m}^2$, comparable to
the maximum obtainable for a {\it normal} neutron star (although requiring a
smaller breaking strain). Converting this to an approximate ellipticity,
assuming the canonical moment of inertia, would give $\varepsilon \approx
2\ee{-5}$.

\subsubsection{Solid strange stars}
In \citet{Xu:2003} the possibility is presented that neutron stars could be made
of solid strange quark matter. \citet{Xu:2003} uses the observation of kHz
quasi-periodic oscillations (QPOs) in X-ray bursts from neutron stars in X-ray
binaries, and their potential association with torsional modes in the star, to
give a shear modulus for solid strange stars of $\mu \approx 4\ee{31}\,{\rm
J}\,{\rm m}^{-3}$. As pointed out by \citet{Owen:2005} the identification of the
QPO frequencies with torsional modes is somewhat problematic, but similarly we
will take this shear modulus to be an upper limit for strange stars. Also
\citet{Lin:2007} notes that the theoretical arguments for solid strange stars
are less robust that those for the crystalline colour superconducting stars
discussed in \S{\ref{sec:css}. The masses and radii for models of strange quark
stars can be seen, for example, in Fig. 2 of \citet{Lattimer:2001}. Although
these may not necessarily hold for {\it solid} strange stars we will use the
range of masses from 0.5--2\,$M_{\odot}$ and radii from 8--11\,km given by this
figure to estimate the likely range of quadrupoles from Eqn.~\ref{eq:quadmom}.
Unlike a {\it normal} neutron star models of quark stars show a decrease in mass
with a decrease in radius. Using the maximum mass and radius and a breaking
strain of $\sigma = 10^{-2}$, gives the highest quadrupole of $Q_{\rm max}
\approx 3.5\ee{34}\,{\rm kg}\,{\rm m}^2$. Converting this to an approximate
ellipticity, assuming the canonical moment of inertia, would give $\varepsilon
\approx 5\ee{-4}$.

\section{Limits from detection}\label{sec:limits}
The spin-down limit assumes that the total spin-down luminosity of the pulsar
is emitted as gravitational radiation, but if we assume that the currently known
pulsars are only {\it just} observable at an S/N of 5 we can estimate what 
{\it angle averaged} percentage of the spin-down power is going into the \gw 
emission. This is shown, for ALV, ET-B and ET-C in
Fig.~\ref{fig:percentofSDlimit}. We can see that for the majority of pulsars for
which we could beat the spin-down limit more than 1 per cent of the spin-down
power would need to be lost via \gws for us to just observe them (with only 12
per cent, 26 per cent and 34 per cent requiring less than this for ALV, ET-B and
ET-C respectively). Less than 1 per cent would be consistent with the one pulsar
for which we have a reliable limit on this for (the Crab pulsar) that is losing
less than 2 per cent of its power in this way \citep{Abbott:2010a}. However, for
the majority of pulsars we have no reliable way to estimate how much energy is
lost via gravitational waves, especially given that only a few {\it young}
pulsars (and no millisecond pulsars) have measured braking indices, so 10s of 
per cent are not ruled out.
\begin{figure}
\includegraphics[width=84mm]{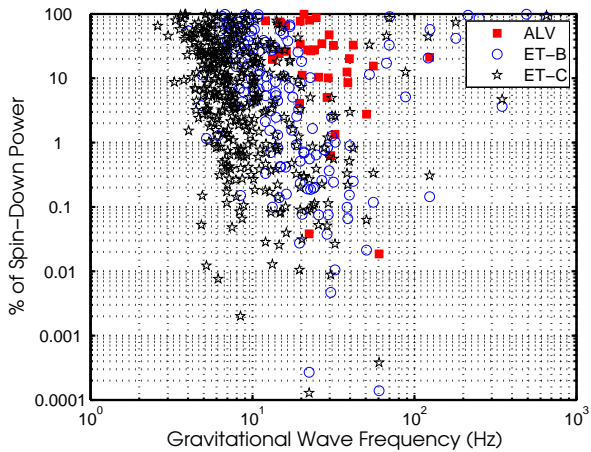}
\caption{The percentage of the spin-down power of known pulsars emitted via
gravitational waves if the pulsar is just observable with an S/N of 5. Squares
represent observation with ALV, circles represent ET-B and stars represent
ET-C.}
\label{fig:percentofSDlimit}
\end{figure}

As we saw in \S\ref{sec:eos} there are large uncertainties in many of the
parameters that give rise to a neutron stars quadrupole moment. The stars could
also just be intrinsically quite un-strained i.e.\ $\langle\sigma_{22}\rangle$
may just be small. This, and the fact that many parameters are highly
correlated, means that \gw observations from the quadrupole of a triaxial
neutron star will generally be unable to pin-down much of the physics giving
rise to it. However, as we have seen in \S\ref{sec:eos} there are reasonably
hard upper limits on the emission allowable for stars with different \eoss, so
if we see emission above a particular \eos limit then these observations could
rule that out for a particular pulsar, or alternatively rule others in.

With this in mind we will look at all currently known pulsars within the
expected sensitive bands of ALV and the Einstein Telescope (in ET-B and ET-C
configurations) and say whether they could be constrained to a particular \eos,
or whether a variety of \eoss are valid. We will do this for pulsars that the
spin-down limit suggests are observable with an S/N of 5 or greater. However, 
it should be noted that our list of \eoss is not exhaustive, and other 
possibilities may exist (whether they have already been theorised or not), so
here constraining a pulsar to a particular \eos may not be the end of the story.

The quadrupole moments for each pulsar that could be observed with ALV, ET-B and
ET-C are shown in Fig.~\ref{fig:Q22s}, both for the case that the pulsar is
emitting at its spin-down limit and also an angle averaged scaled value that
assumes the pulsar is only {\it just} observable at an S/N of 5 (the scaling
factor for each pulsar can be obtained by taking the square root of the values
from Fig.~\ref{fig:percentofSDlimit}). Maximum values for the different \eos
taken from \S\ref{sec:eos} are also shown. The values are calculated from known
spin-down limits, which are not available for globular cluster (GC) pulsars, so
for these we have limits based on two assumptions. The first is a
spin-down-based limit that assumes (see \S\ref{sec:gcs}) that the star has a
maximal \gw spin-down of $\dot{\nu} = -5\ee{-13}$\,Hz\,s$^{-1}$, and the second
is, as above, assuming that the star is observable at an S/N of 5. As shown in
\S\ref{sec:gcs} the values of $h_0$ needed to produce the levels of emission
required for this S/N can be converted into equivalent \gw spin-down values. For
ALV about half, and for ET-B/C all bar one of the pulsars (see
Fig.~\ref{fig:gcsd}), are well within a range that could be masked by
intra-cluster accelerations (we exclude the others). The number of pulsars with
angle averaged quadrupoles below the spin-down limits for ALV, ET-B and ET-C are
59, 295 and 624 respectively (or 77, 408 and 774 respectively for the best case
orientation, and 50, 408 and 531 for the worst case orientation). Note that due
to the reasons discussed in \S\ref{sec:assess} these numbers are not quite the
same as those suggested from an angle averaged S/N.

From Fig.~\ref{fig:Q22s} (summarised in histogram form in
Fig.~\ref{fig:Q_22_hist}) we see that for the vast majority of young pulsars
(generally those below $\sim 40$\,Hz) the S/N limits correspond to large
quadrupoles only supportable by the most extreme \eoss mainly
through the the application of a large breaking strain. We unfortunately cannot
expect the majority of these quadrupoles to be realistic unless there is a
very unexpected large population of highly deformed stars. However, this could
be realistic for a small subset of these pulsars, or some as yet undiscovered
sources. 

It should be stressed that these pulsars could however all be very smooth.
The spin-down limits alone in Fig.~\ref{fig:Q22s} show that the lowest
quadrupoles are only a few $10^{29}$\,kg\,m$^2$ (or converting to ellipticities
a few $10^{-9}$) and all pulsars could be at or below this level. However, even
at this low level Fig.~\ref{fig:Q22s} shows that these may be detectable
at S/N of a few with ET.

\begin{figure}
\begin{tabular}{c}
\includegraphics[width=84mm]{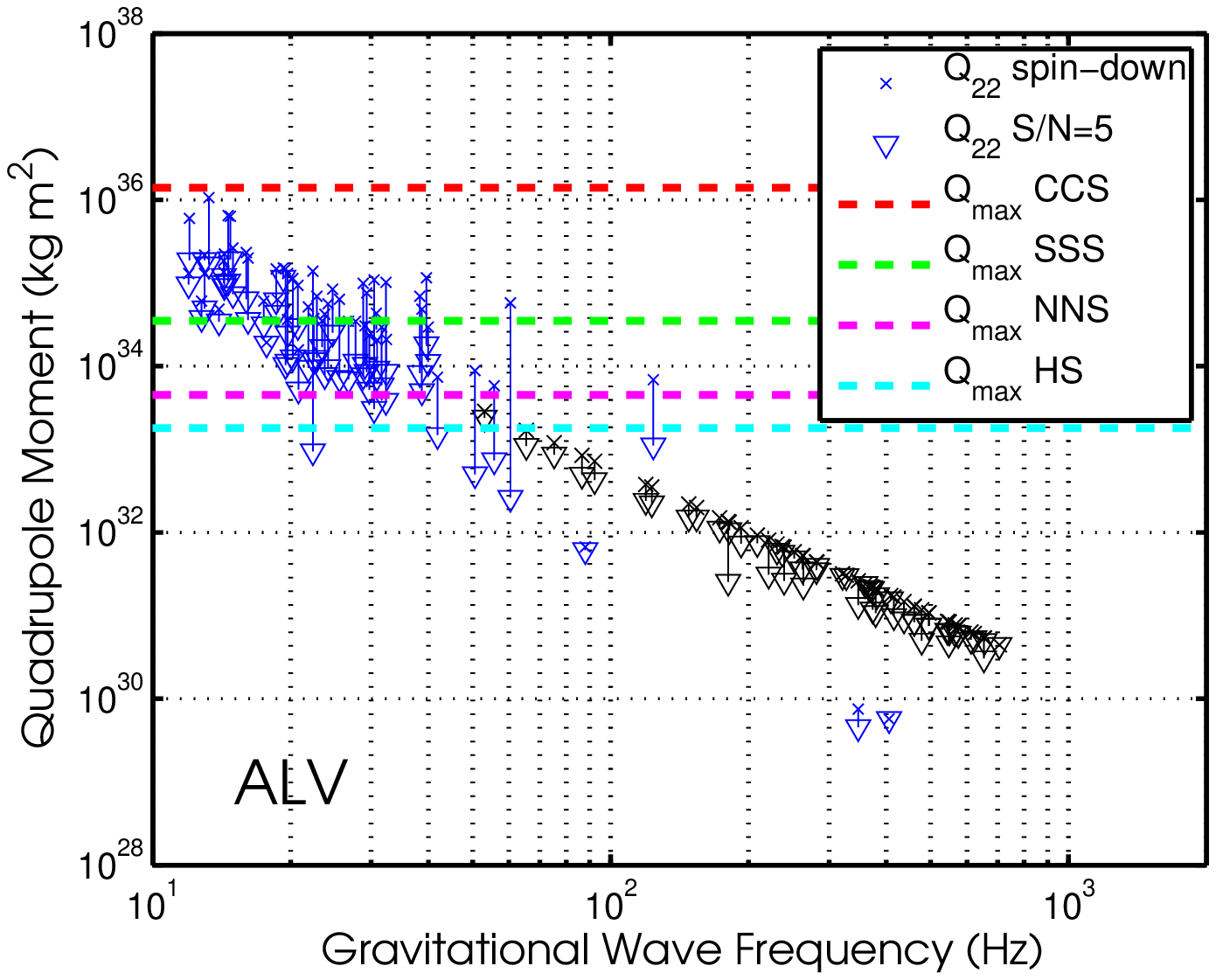} \\
\includegraphics[width=84mm]{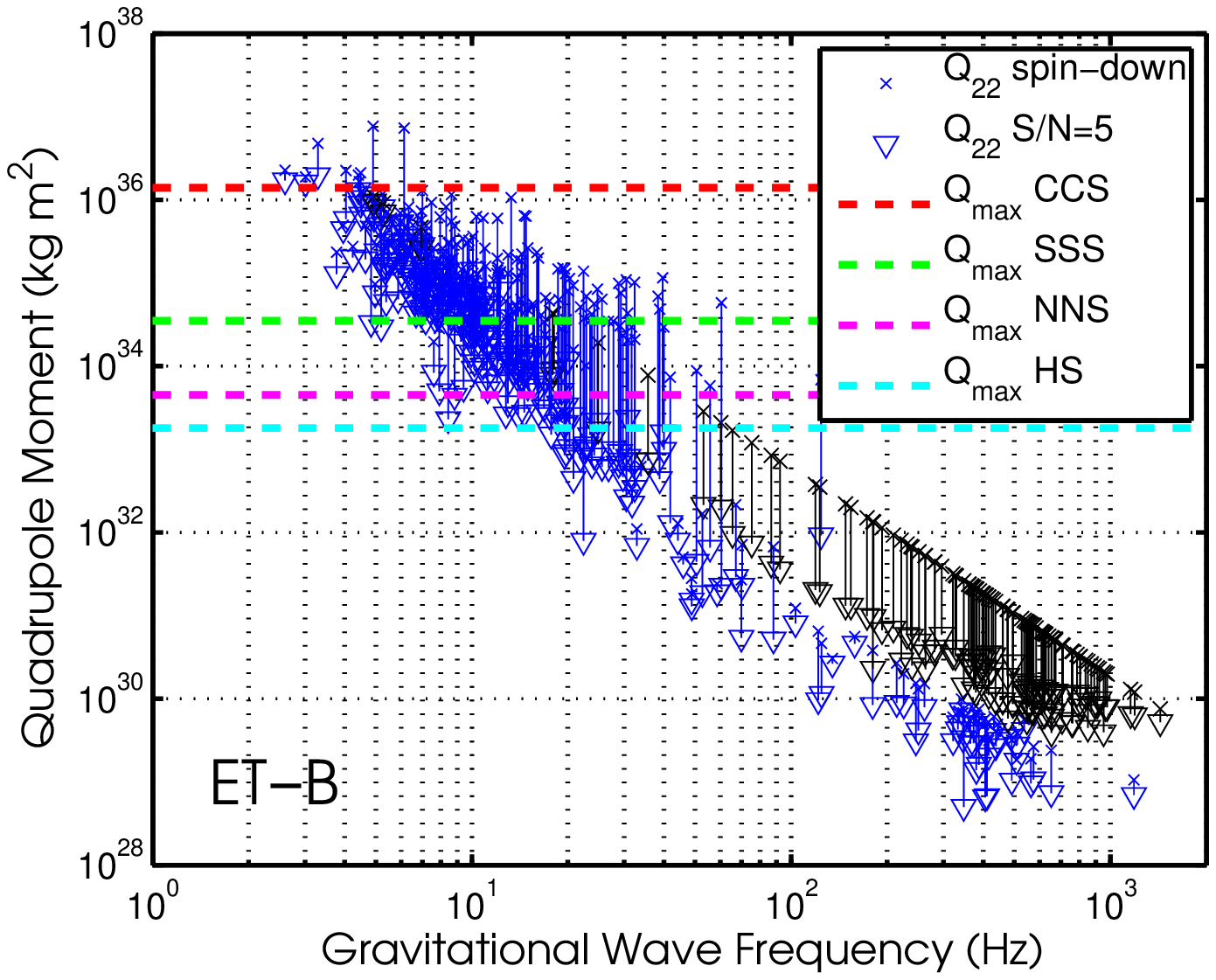} \\
\includegraphics[width=84mm]{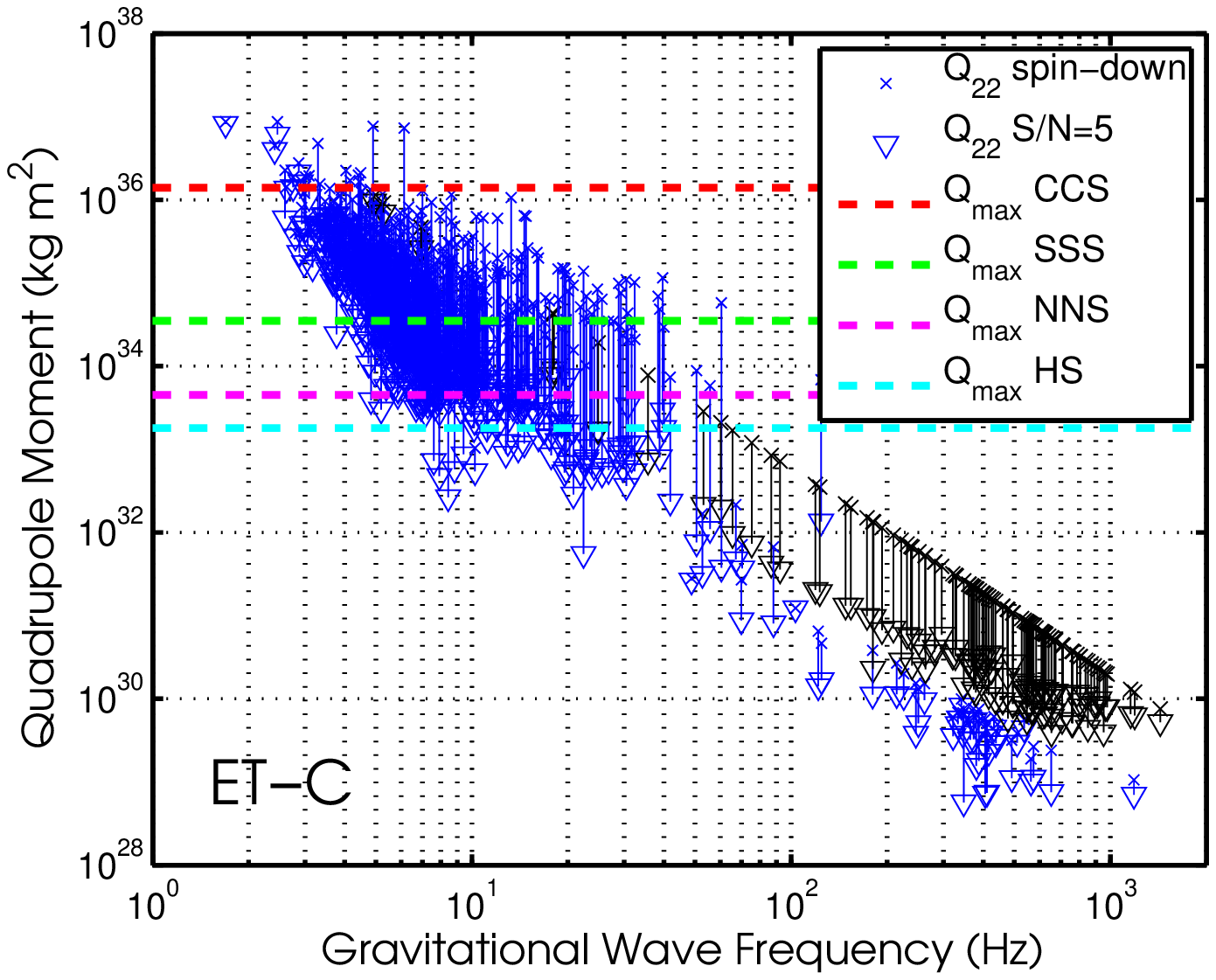}
\end{tabular}
\caption{Necessary quadrupole moments of known pulsars, based on spin-down
limits, for them to be observable with (from top to bottom) ALV, ET-B and ET-C.
Limits based on the observed spin-down, or a spin-down of $\dot{\nu} =
-5\ee{-13}$\,Hz\,s$^{-1}$ for GC pulsars, are given as crosses (blue for non-GC
pulsars and black for GC pulsars). Angle averaged quadrupole values based on
emission at an S/N of 5 are given as triangles. The approximate maximum
quadrupoles from the
various \eoss discussed in \S\ref{sec:eos} are also plotted.}
\label{fig:Q22s}
\end{figure}
\begin{figure}
\begin{tabular}{c}
\includegraphics[width=84mm]{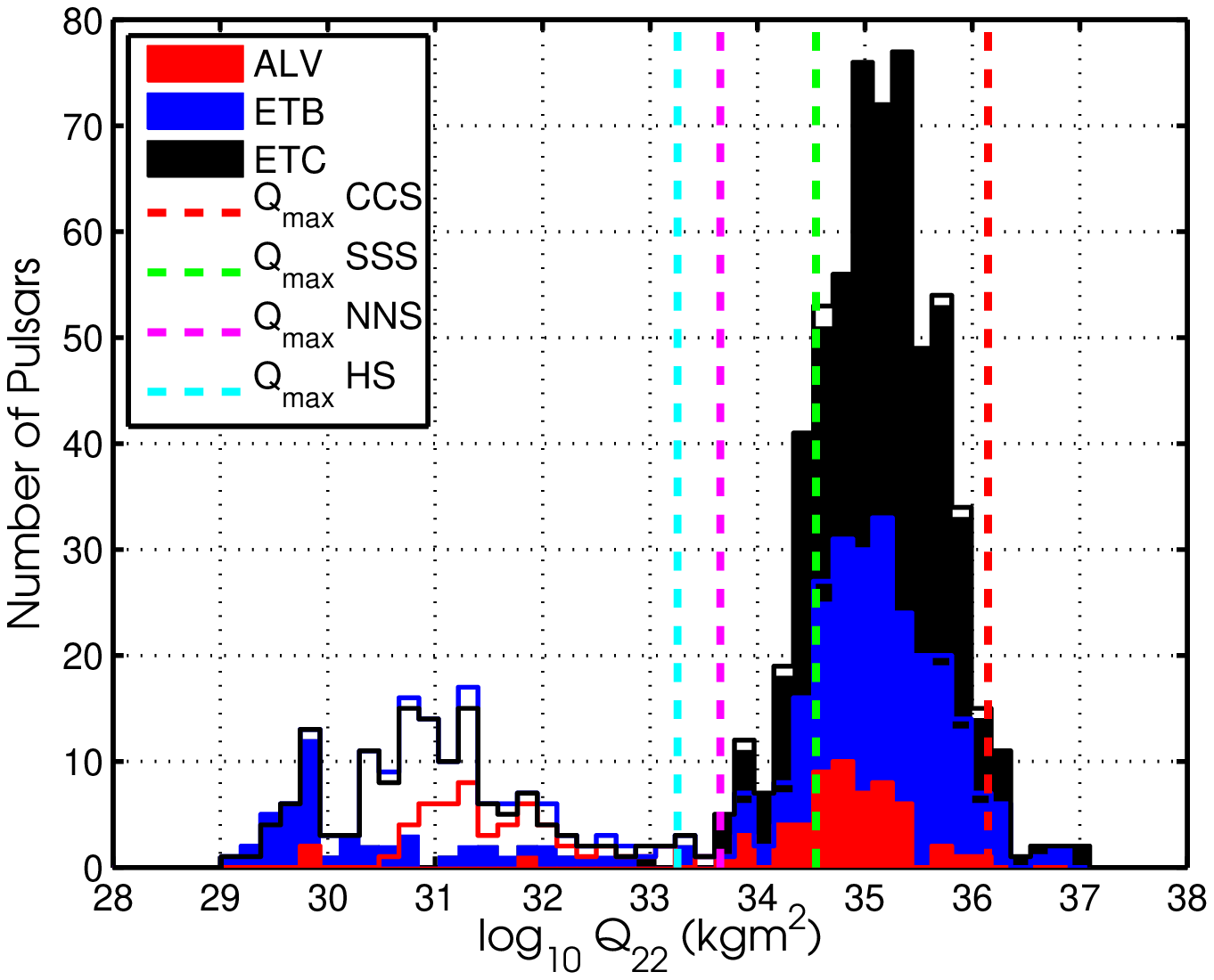} \\
\includegraphics[width=84mm]{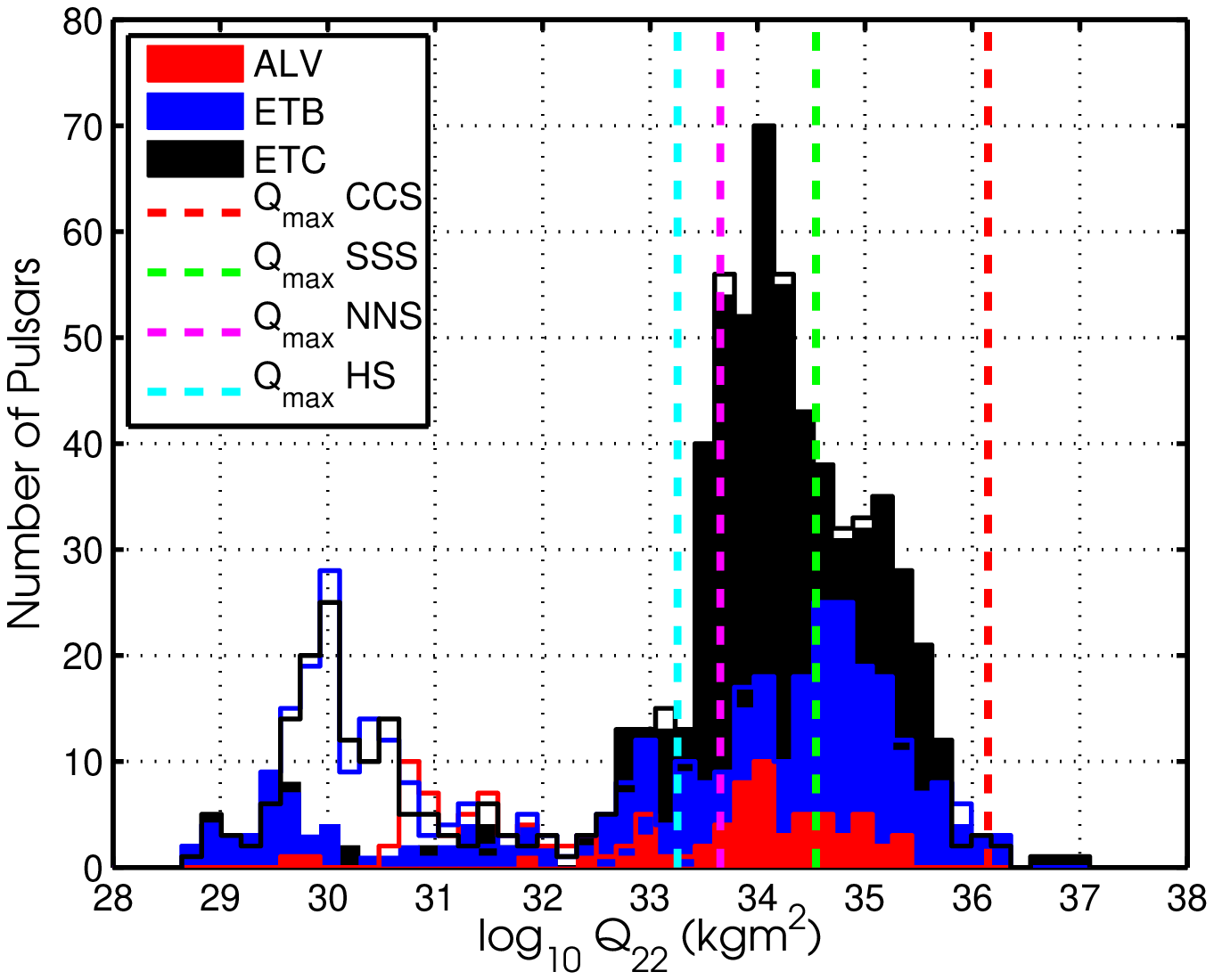}
\end{tabular}
\caption{Quadrupole moments of known pulsars (Fig.~\ref{fig:Q22s}), for them to
be observable with ALV, ET-B and ET-C if emitting at the spin-down limit (top)
and if scaled as to be {\it just} observable at an S/N of 5 (bottom). The filled
histograms show just those pulsars outside of globular clusters and the unfilled
histograms also include pulsars within globular clusters.}
\label{fig:Q_22_hist}
\end{figure}

A summary of the total numbers of pulsars below each \eos limit for pulsars
within, and outside, GCs is given in Table~\ref{tab:summary}.

\subsection{Spin-down limits for Globular Cluster pulsars}\label{sec:gcs}
Several globular clusters have been specifically targeted in radio searches for
millisecond pulsars due to having a large population of old stars. This has lead
to a selection effect giving large numbers of the currently known pulsars
residing in GCs. Pulsars within a GC will undergo significant accelerations,
which contaminate their true spin-down rate $\dot{\nu}_{\rm int}$ via
\begin{equation}
\dot{\nu}_{\rm obs} = \dot{\nu}_{\rm int} - \frac{a_{||}}{c}\nu
\end{equation}
where $\dot{\nu}_{\rm obs}$ is the observed spin-down, $\nu$ is the pulsar
frequency and $a_{||}$ is the pulsar's acceleration along the line-of-site -- in
fact many are observed to have a spin-up due to the accelerations (as they are
millisecond pulsars they have small intrinsic spin-downs anyway, so it is quite
easy for the accelerations to swamp this value). We can therefore use \gw
observations of this set of pulsars, either through upper limits or detections,
to set limits on the \gw component of the intrinsic spin-down rate via
\begin{equation}
\dot{\nu}_{\rm GW} = - h_0^2 \left(\frac{2c^3r^2\nu}{5GI_{zz}}\right).
\end{equation}
We cannot however limit the overall spin-down as, unless the stars acceleration
can be independently assessed and the observed spin-down calculated, there is no
way to know what fraction of the total spin-down is due to \gw emission. As
stated in \citet{Owen:2006} by looking at the observed spin-downs given for GC
pulsars in the Australia Telescope National Facility (ATNF) Pulsar Catalogue
\citep{Manchester:2005} it would be hard for cluster dynamics to mask spin-downs
larger than $\dot{\nu} \sim -5\ee{-13}\,{\rm Hz}\,{\rm s}^{-1}$ (this being the
largest observed spin-up seen for any GC pulsar), so if \gw observations can
limit values to smaller than this, then results could be providing new
information.

In Fig.~\ref{fig:gcsd} the values of the \gw spin-down that the GC pulsars
would require to be observed at S/N of 5 are given. For ALV 52 of the 103
observable GC pulsars, and for ET-B and ET-C 1 of the 107 observable pulsars,
would need spin-downs greater than $-5\ee{-13}\,{\rm Hz}\,{\rm s}^{-1}$,
i.e.\ values that probably could not be masked by accelerations and therefore
would already be seen as having large spin-downs from radio observations. For
the rest of the pulsars the spin-downs they would need could relatively easily
be masked by accelerations, so they are definitely worthwhile as targets for \gw
searches.

\begin{figure}
\begin{tabular}{c}
\includegraphics[width=84mm]{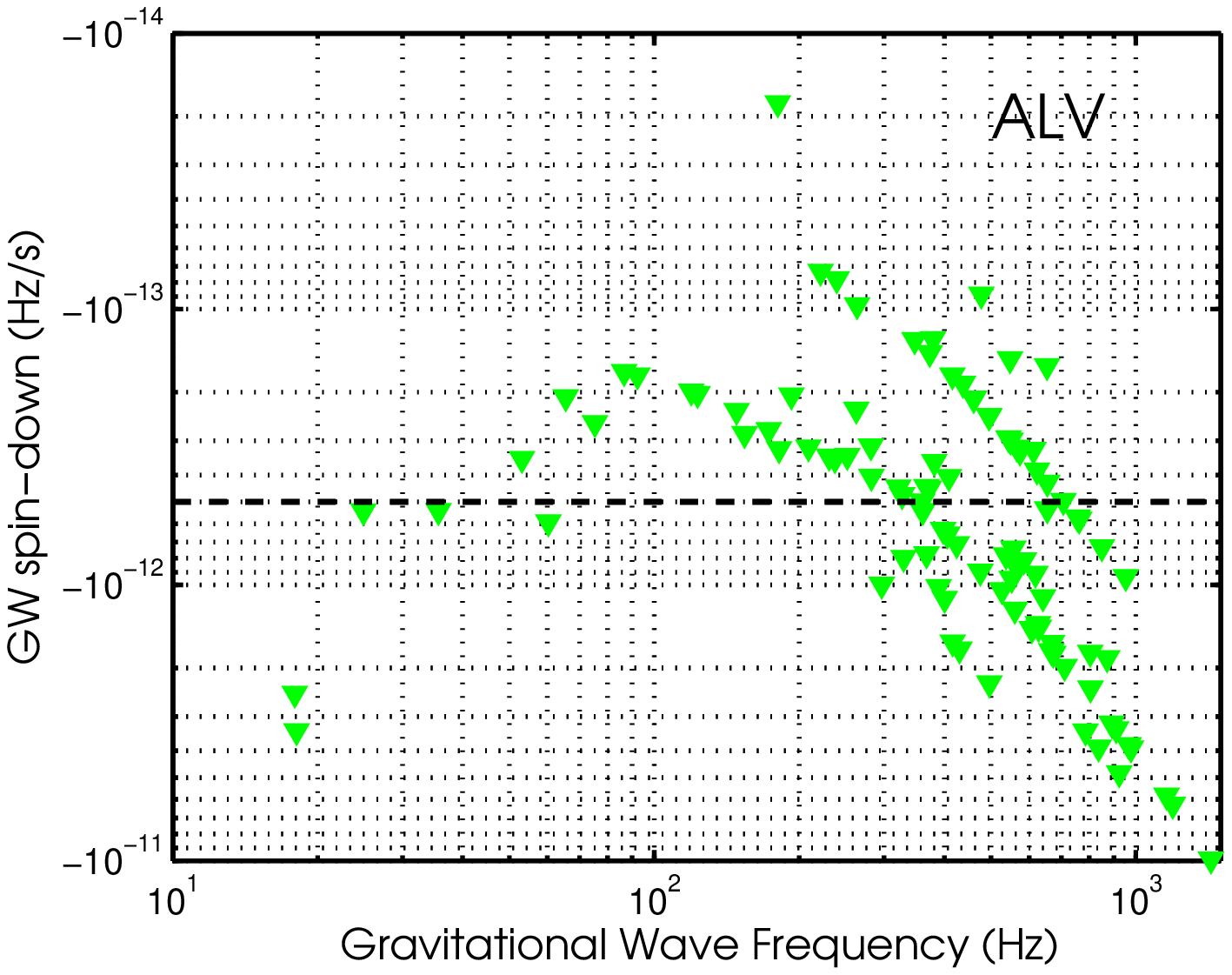} \\
\includegraphics[width=84mm]{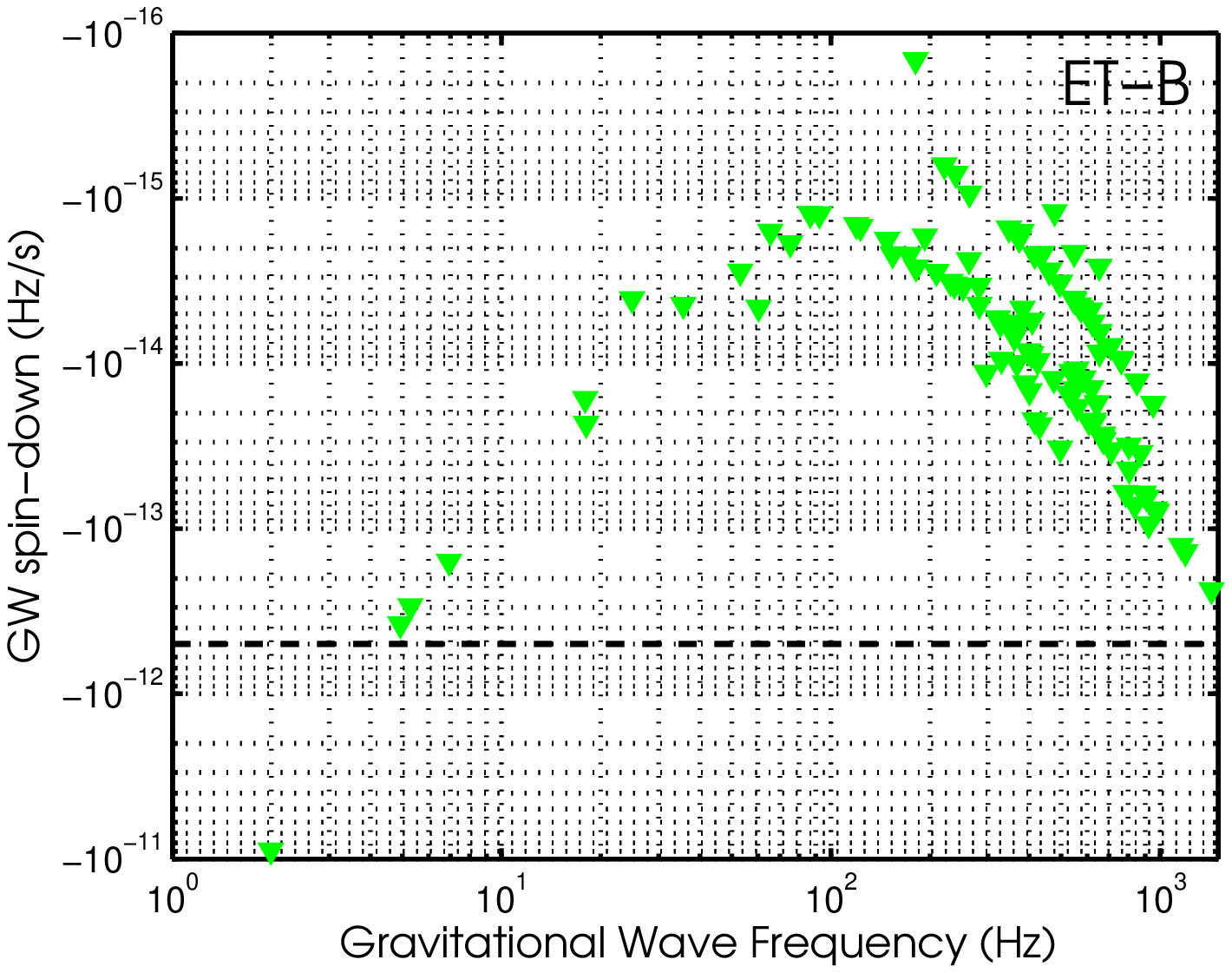} \\
\includegraphics[width=84mm]{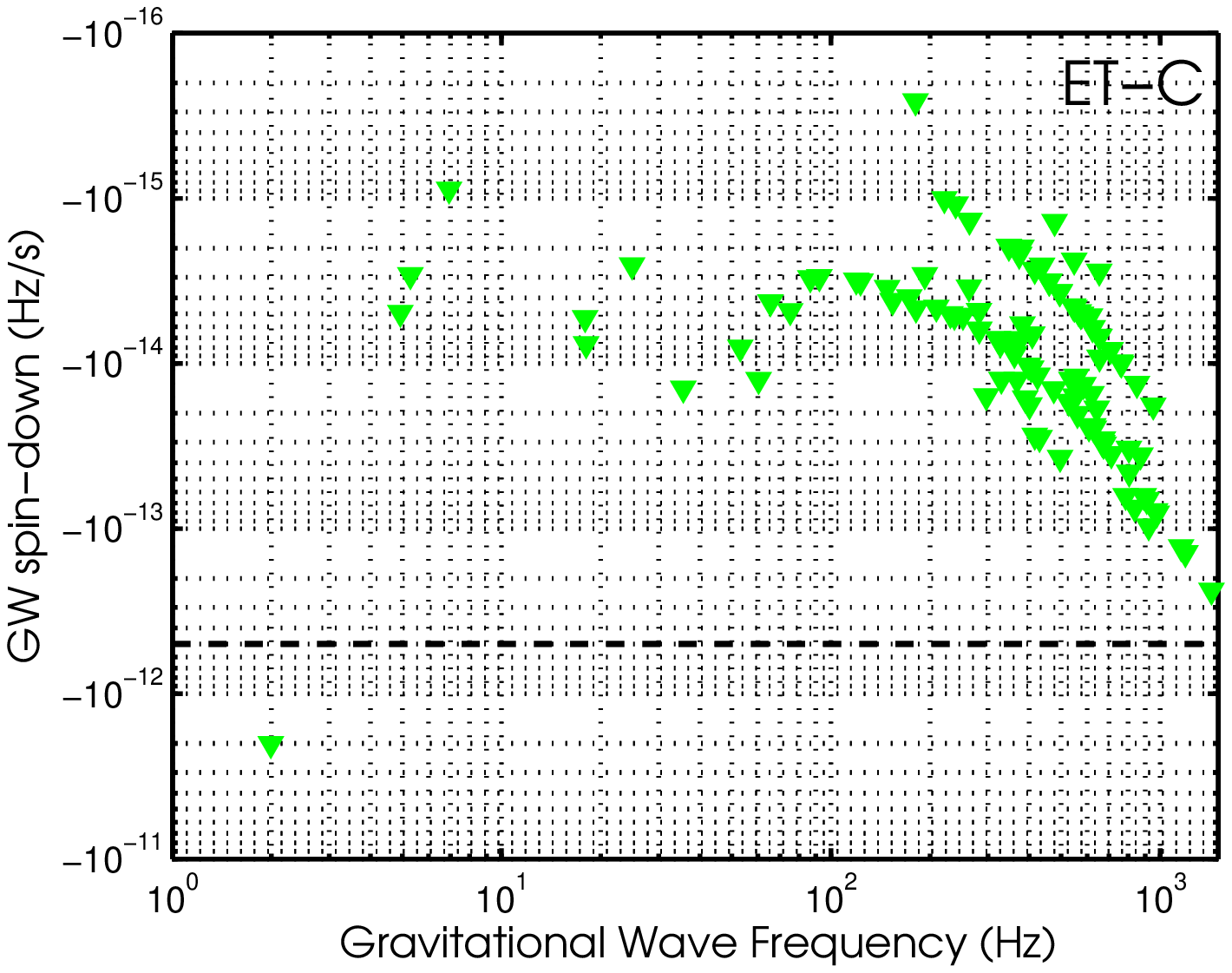}
\end{tabular}
\caption{Spin-downs that would be needed in currently known GC pulsars for them
to be observed with (from top to bottom) ALV, ET-B and ET-C
at an S/N of 5.}
\label{fig:gcsd}
\end{figure}

These limits are the smallest values that the spin-down could have to be
observable, but for many of these pulsars there is a 1--2 orders of magnitude
range between these limits and the (observationally motivated, but slightly
arbitrary) maximum limit of $\dot{\nu} \approx -5\ee{-13}\,{\rm Hz}\,{\rm
s}^{-1}$. So, there is quite a lot of leeway for these pulsars to be emitting
at greater levels than that producing an S/N of 5. Also, it can be seen from
Fig.~\ref{fig:Q22s} that for the majority of these pulsars, both the maximum and
minimum observable quadrupoles are in ranges allowable by all equations of
state.

\begin{table*}
\begin{minipage}{145mm}
\caption{Summary of the quadrupole moment limits given in Figs.~\ref{fig:Q22s}
and \ref{fig:Q_22_hist}. For comparison the total number of non-GC pulsars with
angle averaged quadrupoles below the spin-down limit and observable with an S/N
$> 5$ are 59, 295 and 624 for ALV, ET-B and ET-C respectively. The total number
of GC pulsars potentially observable with S/N $> 5$ and a spin-down limit
smaller than $-5\ee{-13}\,{\rm Hz}\,{\rm s}^{-1}$ is 51 and 106 for ALV, and
ET-B/C respectively.}
\label{tab:summary}
\begin{tabular}{l l c c c c}
\hline
~  & ~ & \multicolumn{4}{c}{Number of pulsars below limit} \\
\hline
~ & ~ & $Q_{\rm max}$ CCS & $Q_{\rm max}$ SSS & $Q_{\rm max}$ NNS & $Q_{\rm
max}$ HS \\
\hline
\multirow{4}{*}{ALV} & spin-down limit & 59 & 17 & 3 & 3 \\
& S/N=5 & 59 & 42 & 11 & 9 \\
& GC spin-down limit ($\dot{\nu}=-5\ee{-13}\,{\rm Hz}\,{\rm s}^{-1}$) & 51 &
51 & 51 & 50 \\
& GC S/N=5 & 51 & 51 & 51 & 50 \\
\hline
\multirow{4}{*}{ET-B} & spin-down limit & 286 & 90 & 53 & 51 \\
& S/N=5 & 292 & 175 & 103 & 85 \\
& GC spin-down limit ($\dot{\nu}=-5\ee{-13}\,{\rm Hz}\,{\rm s}^{-1}$) & 106
& 101 & 99 & 97 \\
& GC S/N=5 & 106 & 103 & 101 & 101 \\
\hline
\multirow{4}{*}{ET-C} & spin-down limit & 605 & 131 & 43 & 41 \\
& S/N=5 & 618 & 421 & 152 & 93 \\
& GC spin-down limit ($\dot{\nu}=-5\ee{-13}\,{\rm Hz}\,{\rm s}^{-1}$) & 106
& 101 & 99 & 97 \\
& GC S/N=5 & 106 & 104 & 102 & 101 \\
\hline
\end{tabular}
\end{minipage}
\end{table*}

\subsection{Limits on magnetic fields}\label{sec:bfield}
Deformations of a neutron star can be supported purely by the strain that the
star can sustain\footnote{How long such a deformation could be sustained due
to visco-elastic creep smoothing it out is something that needs further
study, e.g.\ \citet{Chugunov:2010} who suggest short timescales of a few years
for hot stars, but far longer for cooler stars (although they note
extrapolations to lower temperatures and longer timescales must be treated with
care).}, although the mechanism giving rise to this strain may be unknown.
However, magnetic fields give both a mechanism for producing strains, and a way
of sustaining that strain. The external magnetic fields of most pulsars can be
estimated by assuming that all spin-down is due to magnetic dipole radiation,
and these show that for millisecond pulsars the magnetic field strengths are
comparatively small ($10^8$ to $10^9$ gauss) and nowhere near enough to sustain
an appreciable deformation on the star. The young pulsars have external dipole
fields about 1000 times larger at $\sim10^{12}$ gauss, but this is generally
still not enough to give deformations that would produce observable
gravitational waves. However, these stars could potentially have internal fields
far larger than the external dipole, which would be large enough to give rise to
\gw producing distortions. As done in \citet{Abbott:2010a} we can use \gw
observations, or upper limits, to place limits on this internal field strength
for all the currently known pulsars. \citet{Cutler:2002} predicts that in
{\it normal} neutron stars toroidal magnetic fields could give rise to prolate
stars with ellipticities of order
\begin{equation}
\varepsilon = 1.6\ee{-6} \times 
\begin{cases}
\langle B_{15} \rangle {\rm G}, & B<10^{15}\,{\rm G}, \\
\langle B_{15}^2 \rangle/ {\rm G}^2, & B>10^{15}\,{\rm G},
\end{cases}
\end{equation}
where $\langle B_{15} \rangle$ is the volume averaged magnetic field in units of
$10^{15}$\,G. \citet{Haskell:2008} also study the role of internal magnetic
fields, both entirely poloidal and toroidal, and how this would effect the
star's ellipticity (in particular for a star with an \eos described by an $n=1$
polytrope). They give
\begin{equation}
\varepsilon \approx R_{10}^4 M_{1.4}^{-2} \langle B_{15}^2 \rangle \times
\begin{cases}
2\ee{-4},~{\rm poloidal}, \\
-1\ee{-6},~{\rm toroidal},
\end{cases}
\end{equation}
which for the toroidal case is similar to that of \citet{Cutler:2002}. Very
similar limits for toroidal fields in superconducting stars are given by
\citet{Akgun:2008}. These equations can be re-arranged to give limits on the
magnetic fields given \gw observations of the quadrupole (see \S\ref{sec:eos})
and an assumed moment of inertia, which here we will take as
$10^{38}$\,kg\,m$^2$ (as noted above this is probably a lower limit and could
differ by up to a factor of $\sim 3$). Potential measurements of the field
strength for both the poloidal and toroidal cases if signals were observed at an
S/N of 5 for the ALV, ET-B and ET-C set ups are given in Fig.~\ref{fig:bfield}.

\begin{figure}
\begin{tabular}{c}
\includegraphics[width=84mm]{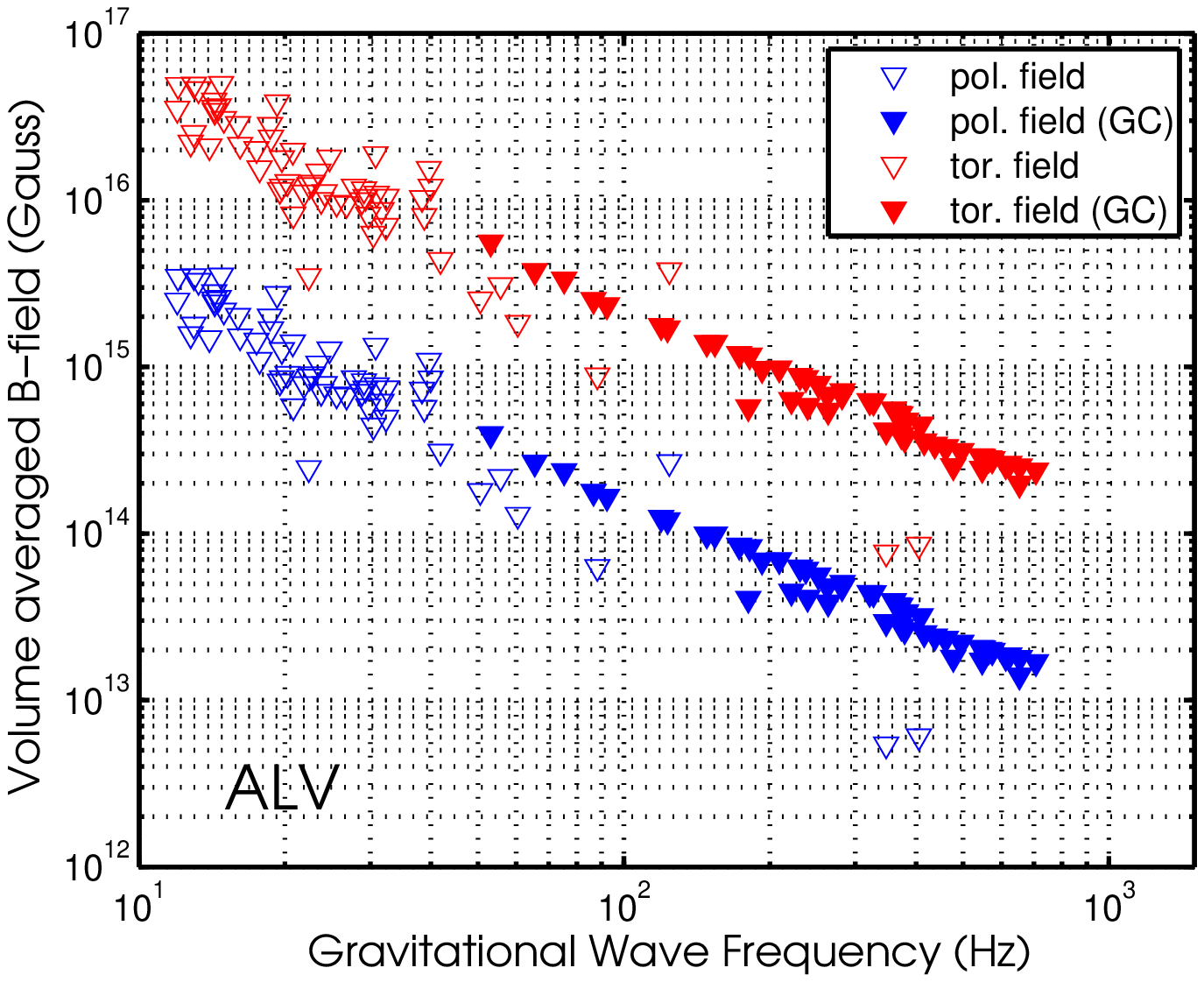} \\
\includegraphics[width=84mm]{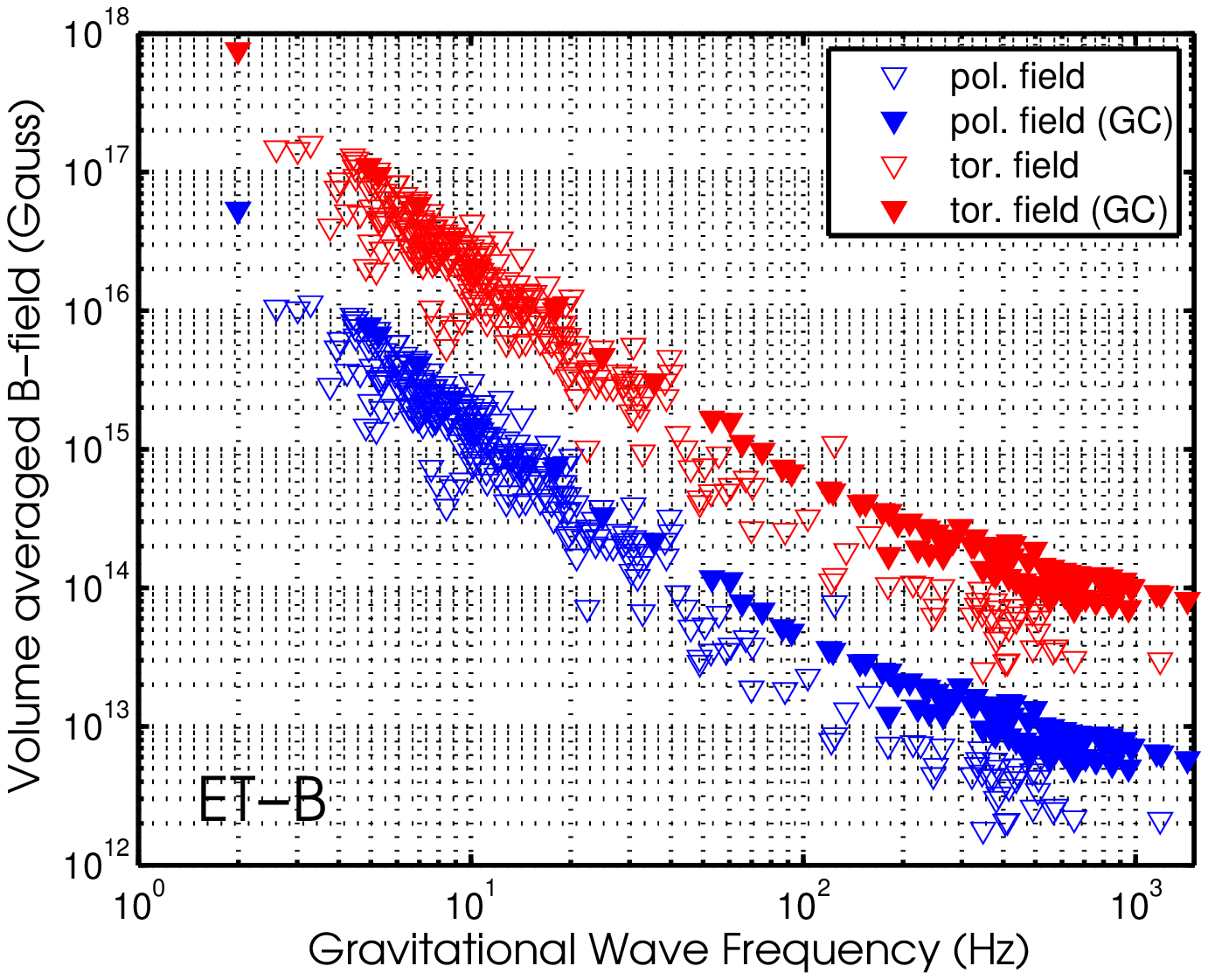} \\
\includegraphics[width=84mm]{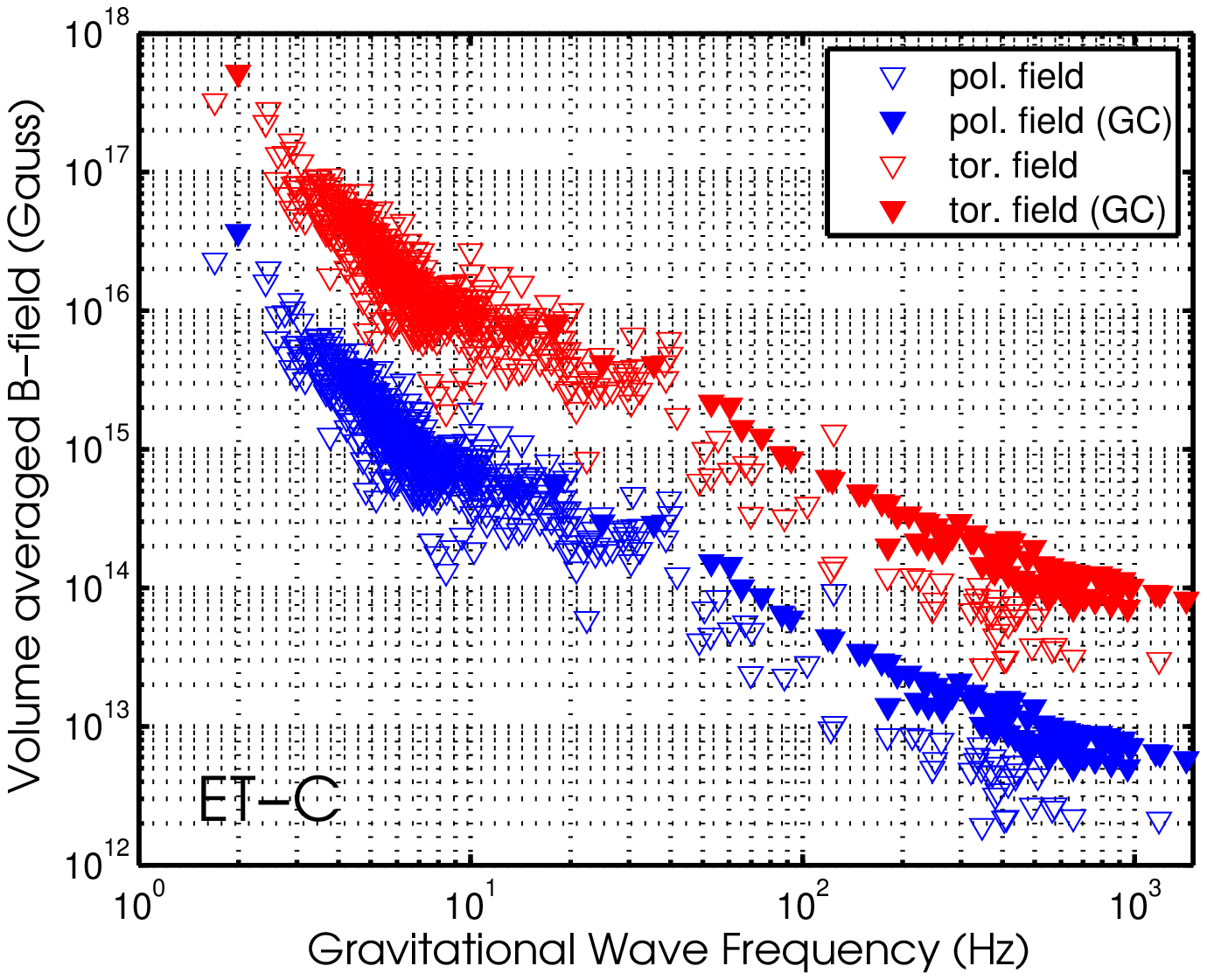}
\end{tabular}
\caption{Potential internal magnetic field strengths that would be needed in
currently known pulsars to observed them with an S/N of 5 with (from top to
bottom) ALV, ET-B and ET-C.}
\label{fig:bfield}
\end{figure}

We can see that for the young pulsars (with frequencies $\lesssim 40$\,Hz)
internal magnetic fields of greater than $10^{14}$\,G would be required to
observe them, which is of a similar strength to the {\it external} fields of a
magnetars at $10^{14}-10^{15}$\,G. For the millisecond pulsars, in particular
the non-GC pulsars, internal poloidal fields of a few $10^{12}$\,G could
provide observable signals, although if there are only toroidal fields then a
couple of orders of magnitude higher would be necessary. Of course the field
geometry could be complex and a combination of toroidal and poloidal
components, and the calculations above rely on a specific \eos.

\section{Discussion}
We have looked at our ability to detect and estimate parameters from \gw
observations for known (non-accreting) pulsars. Using Bayesian
hypothesis testing, we estimate that signals with an S/N of 5 could be detected
with 95 per cent efficiency. This assumes that the data is Gaussian, and
experience of
real detector data shows that this assumption can be reasonable on short time
scales and with effectively cleaned data, although there will still probably be
some small degrading of detection ability.

Once detected, we have shown how estimates of various parameters will be
affected as signals increase in S/N, and also how the pulsar's orientation
affects this. We see that for strong signals in the case where the orientation
is most favourable and gives a larger S/N (i.e.\ the \gws are circularly
polarised) the uncertainty on the \gw amplitude will actually be larger than an
equivalent amplitude source with a worse orientation. So, when detections become
regular the best parameter constraints will actually be made for the linearly
polarised sources. However, we also see that unless the distance to these
pulsars can be measured to better than $\sim$10 per cent then the fractional
error on
the quadrupole calculated from the \gw amplitude will be relatively insensitive
to the orientation, because the uncertainty is dominated by the distance
uncertainty.

We have seen the sorts of mass quadrupoles that would be necessary to observe
the set of currently known pulsars with future \gw observatories, and compared
these to maximum theoretical predictions for a variety of \eoss.
As has previously been noted many times (e.g.\ \citealp{Abbott:2007a}) for the
majority of millisecond pulsars, if we are able to beat their spin-down limits
and observe them, the quadrupoles (or ellipticities) they would have are
sustainable by all \eoss and would therefore not be able to
constrain the type of matter in the star. However, with observations of many
pulsars useful population statistics could be obtained and potential differences 
between different populations explored (e.g.\ GC pulsar and non-GC pulsars). To 
be observed with ALV, ET-B or ET-C the majority of young pulsars (with 
frequencies $\lesssim 40$\,Hz) would have to have quadrupoles greater than 
$\approx 10^{33}$\,kg\,m$^2$, which would only be sustainable for highly strained
crystalline colour-superconducting stars or solid strange quark stars. 
Therefore if such stars were observed it would provide a great deal of insight
into neutron star matter. The number of pulsars potentially observable at an S/N
of 5, and
consistent with being a {\it normal} neutron star is 11 for ALV, 103 for ET-B and 152
for ET-C. It is important to note that this does not mean that they {\it will}
have these quadrupoles and they could all be smoother than our limits are able
to set. 

In \citet{Andersson:2009}, which studies the prospects of observing \gws
from neutron stars with ET (in particular the ET-B configuration), they take the
assumption that realistic stars would optimistically have ellipticities
$<10^{-7}$ and conclude that only some of the 100s of millisecond pulsars
currently known would be potentially observable. From this work we can look at 
the number of known pulsars we might expect to be observable at an S/N greater
than 5
given that they are losing less than a specific percentage of their spin-down
power through \gw emission, and have quadrupoles below specific levels (this is
shown in Table~\ref{tab:overall}). We find that for ALV if we take pulsars that
would only need to be emitting less than 50 per cent of their spin-down
luminosity via
\gws to be observable at S/N of greater than 5, and had quadrupoles less than
$Q_{22} < 10^{30}$--$10^{31}$\,kg\,m$^2$ (or ellipticities of $\lesssim
10^{-8}$--$10^{-7}$), then only three of the currently known pulsars would be
observable. We find that none are observable if requiring these quadrupole
levels and also emitting less than 10 per cent of the spin-down power via \gws.
For ET
(in either ET-B or ET-C configurations) if we take pulsars which would only need
to be emitting less than 10 per cent of their spin-down luminosity, and with
quadrupoles $< 10^{30}$--$10^{31}$\,kg\,m$^2$, then 6--11 currently known
pulsars would be observable respectively. For the one pulsar that we can
currently calculate a limit on the spin-down power we see that less than 2 per
cent of
the spin-down luminosity is emitted via gravitational waves, so using this as a
guide and taking only pulsars that could emit less than 1 per cent of their
spin-down
power in \gws and still be observable (and with the above quadrupoles) we find
only one pulsar. These numbers are based on currently known pulsars and use only
the non-GC pulsars; for the millisecond pulsars that these results represent
(due to the small quadrupole requirements) approximately half are within GCs, so
assuming there is no difference in spin-downs between GC and non-GC populations
of pulsars then the number of observable pulsars could double. Also, as stated
in \S\ref{sec:snrestimates}, future radio telescopes such as the SKA could
observe $\approx 1000$ millisecond pulsars increasing the current number by a
factor of 5. This could give 10s of observable pulsars for ET.

\begin{table}
\caption{The number of pulsars potentially observable given gravitational wave 
emission consuming less than various percentages of the total spin-down power,
and requiring quadrupoles less than the given values.}\label{tab:overall}
\begin{tabular}{l c c c c}
 & \multicolumn{4}{c}{Number of pulsars} \\
\hline
$Q_{22}$\,kg\,m$^{2} < $ & $10^{33}$ & $10^{32}$ & $10^{31}$ & $10^{30}$ \\
$\varepsilon \lesssim$ & $10^{-5}$ & $10^{-6}$ & $10^{-7}$ & $10^{-8}$ \\
\hline
Per cent of spin-down power & \multicolumn{4}{c}{$<100$ per cent} \\
\hline
ALV & 8 & 5 & 4 & 1 \\
ET-B & 51 & 45 & 37 & 27 \\
ET-C & 40 & 38 & 33 & 25 \\
\hline
Per cent of spin-down power & \multicolumn{4}{c}{$<50$ per cent} \\
\hline
ALV & 5 & 3 & 3 & 1 \\
ET-B & 37 & 32 & 27 & 19 \\
ET-C & 28 & 26 & 23 & 15 \\
\hline
Per cent of spin-down power & \multicolumn{4}{c}{$<10$ per cent} \\
\hline
ALV & * & * & * & * \\
ET-B & 15 & 13 & 11 & 6 \\
ET-C & 12 & 10 & 9 & 6 \\
\hline
Per cent of spin-down power & \multicolumn{4}{c}{$<1$ per cent} \\
\hline
ALV & * & * & * & * \\
ET-B & 2 & 2 & 1 & 1 \\
ET-C & 1 & 1 & 1 & 1 \\
\end{tabular}
\end{table}

We see that for many millisecond pulsars, if they have internal poloidal
magnetic fields similar in strength to the external fields of young pulsars, or
toroidal field similar in strength to the external field of magnetars, then
they may sustain ellipticities that make them observable. In reality the
internal field will probably consist of both poloidal and toroidal components.
Young pulsar would require far higher magnetic fields for them to sustain
ellipticities that would allow them to be observable. 

The above estimates have all assumed a year of observation. However, the future
detectors could run over several years, and the third generation detectors may
even be the premier \gw observatories spanning decades. There could also be other
similarly sensitive detectors added to the network (e.g.\ the LCGT). All of
these would increase the ability to make observations of, and hopefully refine
parameter estimation for, known pulsars, although the scaling of these
improvements (for equivalent detectors) will only be the square root of the
observation time. Better prospects could come from a far larger selection of
pulsar targets from future radio/X-ray/$\gamma$-ray surveys.

One of the main points to note here is that all the above calculations have
significant uncertainties to them. The theoretical study of neutron
star \eoss is still an area of study with large uncertainties. 
Neutron star mass measurements currently seem to favour fairly conventional 
\eoss, and our upper limits for these may be a fair representation 
of reality. However, the current crop of known pulsars may harbour a variety of 
stellar types and our observations may be the only way to differentiate between 
them. More theoretical understanding through simulations of maximum breaking 
strains of quark matter and hybrid star matter would be of great help in making 
these constraints. In addition to determining whether a star is physically
capable of supporting an observably large strain it is of great importance for
future studies to find mechanisms of producing and sustaining such strains.

The study here has also assumed emission from a triaxial neutron star emitting
at precisely twice the observed electromagnetic frequency via the $Q_{22}$ mass
quadrupole. Potentially more interesting physics could be extracted if the \gw
signal and electromagnetic signal were not so closely aligned (e.g.\ if pulsar
timing noise is only present in the electromagnetic signal \citealp{Jones:2004,
Pitkin:2007}). Deviations between the two could uncover information about the
coupling of the \gw producing component and electromagnetic-producing component,
and emission may occur at the rotation rate \citep{Jones:2010}. Also, \gws from
other vibrational modes, such as long-lived {\it r}-modes \citep{Arras:2003,
Owen:2010}, or fundamental modes (possibly excited by glitches), may be able to
tell us far more about the neutron star \eos. 

\section*{Acknowledgements}
This work has been funded under a UK Science and Technology Facilities Council
rolling grant. I am very grateful to many members of the LIGO Scientific
Collaboration and Virgo Collaboration continuous wave search group for useful
discussions and suggestions on the issues in this paper, and for reading through
early drafts. In particular I would like to thank Graham Woan, Ian Jones, Ben
Owen and Keith Riles for their comments. I would also like to thank Roy Smits
for the useful information he provided.

\end{document}